\pdfoutput=1
\documentclass[aps,prb,twocolumn,superscriptaddress]{revtex4-1}
\usepackage{dcolumn}
\usepackage{bm}
\usepackage{amsmath,amssymb,graphicx,color}
\usepackage{hyperref}
\usepackage[english]{babel}

\newcommand{\w} {\ensuremath{\omega}}
\newcommand{\ket}[1] {\ensuremath{|#1 \rangle}}
\newcommand{\sza} {\ensuremath{\sigma_{z1}}}
\newcommand{\szb} {\ensuremath{\sigma_{z2}}}
\newcommand{\ad} {\ensuremath{a^{\dagger}}}
\newcommand{\eps}{\ensuremath{\epsilon}}

\newcommand{\figref}[2]{Fig.~\ref{#1}{\bf{}#2}}
\newcommand{\gam}{\ensuremath{\gamma_{\phi}}}
\hypersetup{bookmarksnumbered=true,	unicode=false,	pdfstartview={FitH}, pdfnewwindow=true, colorlinks=true,linkcolor=blue, citecolor=blue, filecolor=blue, urlcolor=blue}

\begin{document}
\title{Coupling Two Spin Qubits with a High-Impedance Resonator}

\author{S.P. Harvey}
\author{C.G.L. B\o ttcher}
\author{L.A. Orona}
\affiliation{Department of Physics, Harvard University, Cambridge, MA,
	02138, USA}
\author{S. D. Bartlett}
\author{A. C. Doherty}
\affiliation{Centre for Engineered Quantum Systems, School of Physics,
	The University of Sydney, Sydney, NSW 2006, Australia}
\author{A. Yacoby}
\affiliation{Department of Physics, Harvard University, Cambridge, MA,
	02138, USA}

\begin{abstract}
Fast, high-fidelity single and two-qubit gates are essential to building a viable quantum information processor, but achieving both in the same system has proved challenging for spin qubits. We propose and analyze an approach to perform a long-distance two-qubit controlled phase (CPHASE) gate between two singlet-triplet qubits using an electromagnetic resonator to mediate their interaction. The qubits couple longitudinally to the resonator, and by driving the qubits near the resonator's frequency they can be made to acquire a state-dependent geometric phase that leads to a CPHASE gate independent of the initial state of the resonator. Using high impedance resonators enables gate times of order 10 ns while maintaining long coherence times. Simulations show average gate fidelities of over 96\% using currently achievable experimental parameters and over 99\% using state-of-the-art resonator technology. After optimizing the gate fidelity in terms of parameters tuneable in-situ, we find it takes a simple power-law form in terms of the resonator's impedance and quality and the qubits' noise bath.
\end{abstract}
\maketitle{}
\section{Introduction}
Spin qubits with gateable charge-like states have many desirable features for quantum computing, and have been pursued through a range of qubit implementations including singlet-triplet ($S$-$T_0$) and hybrid qubits in a double quantum dot (DQD) as well as exchange-only qubits in triple dots\cite{petta2005a,wu2014, eng2015,kim2014a,medford2013}.  Coupling to charge speeds up many crucial operations, including single and two-qubit operations and measurement, compared with a solely magnetic control, but they retain coherence times that are orders of magnitude above those of pure charge qubits.  For instance, implementations of $S$-$T_0$ qubits in GaAs boast $>98\%$ fidelity single gate operations up to several GHz as well as $98\%$ measurement fidelity in 1 $\mu$s\cite{cerfontaine2016,nichol2017,shulman2012}.  However, the spin-like nature of these qubits typically leads two-qubit gates to be much slower than single-qubit gates and to have speeds that fall off sharply with distance, making scaling to more than two qubits challenging\cite{shulman2012}. One way to remedy both of these issues is to couple two distant qubits using a resonator\cite{viennot2015, mi2018,samkharadze2018,landig2017}. We consider electric coupling between the resonator field and a charge-like state of a spin qubit, focusing on the $S$-$T_0$ qubit, although we note that it is possible to use any of the spin qubits with gateable charge-like states. 

The $S$-$T_0$ qubit's logical subspace consists of the hyperfine-degenerate singlet and triplet states of two electrons in a tunnel coupled DQD. While electrons in the singlet state are hybridized between the two dots in the ground state, with a distribution determined by the dots' relative energies and their tunnel coupling, the electrons in the triplet state are Pauli-blockaded with one electron in each dot. The $S$-$T_0$ energy splitting, $ J$, is thus controlled by the difference in chemical potentials of the two dots, $\eps$, which can be tuned by proximal RF gates on nanosecond timescales. A magnetic field gradient between the dots drives rotations around $\sigma_x$, but will be neglected for the remainder. 

The qubit's electric dipole operator is diagonal in the energy basis, so the qubit-resonator coupling is a longitudinal interaction\cite{blais2004,kerman2013,schuetz2017,billangeon2015,didier2015,richer2016,royer2017}. The resulting gate that we describe in this paper has a number of advantages over transverse-coupled gates. It is not necessary to  bring the qubits into resonance with one another or the resonator, making the gate quite simple; it relies on applying a single tone near the resonator's frequency to each of the qubits, with no direct control of the resonator required. Moreover, this makes it compatible with remaining at sweet spots for enhanced dephasing time. The gate speed is a linear function of the drive, so it can be turned completely off and does not require high powers for fast gates. Furthermore, there is no Purcell effect and no dispersive approximation is necessary, so the drive frequency can be near the resonator's frequency and the drive amplitude is unconstrained, enabling faster gates. Another advantage is that the gate is independent of the resonator's initial state, and only depends on its dynamics (e.g., its decay rate).  As a result, cross-talk to the resonator and elevated temperatures are not barriers to implementing the gate.
 \section{Geometric Phase Gate}
 
We begin by outlining the essential physics underlying the two-qubit gate. By driving $\eps$ of the first qubit near the resonator's frequency, we cause electrons in the singlet state to oscillate between the two quantum dots while electrons in the triplet state remain stationary due to Pauli blockade. The resonator is thereby excited in a qubit-state-dependent manner, which in turn acts to drive $\eps$ in the far qubit. When the two qubits are driven at the same frequency, the interaction with the resonator has a non-zero average, and the qubits accrue a resonator-dependent geometric phase that lets us perform a CPHASE gate.  We can now consider the main noise processes of this interaction. Driving the qubits closer to the resonator's frequency excites the resonator more, which makes the gate faster, but causes more photons to be lost from the resonator.  For similar coupling mechanisms in other systems with a far larger resonator decay rate than qubit dephasing rate, fidelity can be optimized by driving at a frequency that equalizes dephasing through the qubits and the resonator\cite{elman2017a,gambetta2008,schuetz2017}. However, in the case of an $S$-$T_0$ qubit coupled to a superconducting resonator, the resonator decay rate is comparable to the qubit dephasing rate. In this regime of large qubit dephasing, noise from the resonator is relatively unimportant, and it is essential to perform the gate as fast as possible. The maximum fidelity is achieved when the detuning is set so that the CPHASE gate is performed in a small, integral number of oscillations. This approach is known as a geometric phase gate\cite{leibfried2003a,leung2018,roos2008}. 

\begin{figure}
 \includegraphics[scale=0.6]{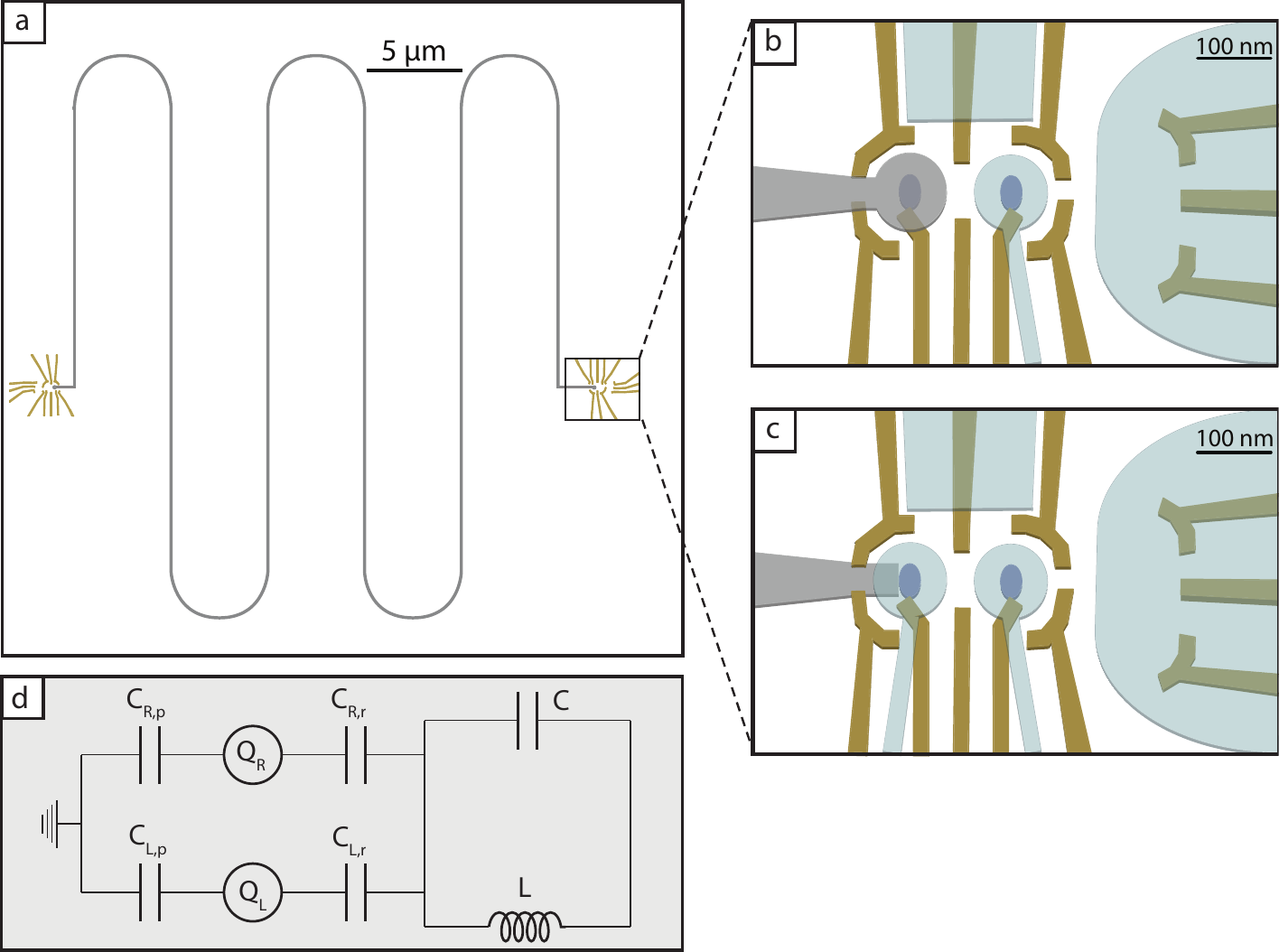}
	\caption{\textbf{a}, Schematic of the two-qubit-resonator device. A double quantum dot is placed at either end of a high-impedance resonator.  A nanowire resonator is shown, but other types have similar dimensions; the resonator can be further meandered to reduce its footprint, or straightened to transport information over larger distances. \textbf{b-c}, Proposed designs for Si-SiGe quantum-dot-resonator devices. Accumulation gates are shaded pale blue, depletion gates gold, and the resonator gray. In \textbf{b}, the resonator replaces one of the accumulation gates, so it is separated by an additional 50 nm oxide, $c_r = 0.02$. In \textbf{c}, the resonator is at the depletion gates layer, $c_r=0.25$. \textbf{d}, A circuit schematic of a qubit resonator system (the left qubit is not shown, but is identical to the right qubit). The left and right quantum dots, $Q_L$ and $Q_R$, have capacitance $C_{i,r}$ to the resonator and total capacitance $C_{i}=C_{i,r}+C_{i,p}$, $i = \{L,R\}$. The resonator has inductance $L$ and capacitance $C$. 
	\label{device}
	}
\end{figure}

We quantify the strength of the qubit-resonator coupling by analyzing the effect of the resonator on the qubit's splitting. The voltage along a high-impedance resonator is much larger than for a conventional 50 $\Omega$ resonator \cite{stockklauser2017,samkharadze2016}; the voltage at the resonator's antinode due to a single photon is $V_0= \sqrt{\hbar Z_r} \w_r $, where $\w_r$ is the resonator's frequency and $Z_r$ its impedance. Quantum-dot-based qubits are compatible with high impedance resonators because they have only tens of attofarads of capacitance and thus have little parasitic effect on the impedance, $Z_r = \sqrt{L/C}$, where $L$ and $C$ are the total inductance and capacitance in the system, respectively.  We consider exciting the resonator near its fundamental frequency, i.e., a half-wave resonator, and place qubits at its antinodes (Fig.~1a). The voltage at each antinode can be written $V_r=V_0 (a + \ad)$.  This voltage shifts the chemical potentials of the quantum dots, which are characterized by a capacitance matrix describing the interactions between each dot and its electrostatic environment \cite{vanderwiel2002}. Denoting the DC contributions to this chemical potential by $\eps_{0}$, we can write $ \eps = \eps_{0} + e c_g V_g + e c_r V_r$, where $e$ is an electron charge, and $c_g$ and $c_r$ represent the lever arms between the double quantum dot and the RF gate and resonator, respectively, and which determine the shift in chemical potential of the DQD caused by a voltage shift on those gates. We define the drive on the RF gate as $e V_g c_g = \eps_d \cos{\w_d t}$.  
We then expand $J$ around $\eps_0$ to second order: 
\begin{multline} \label{eq:J}
  J(\epsilon) \approx J(\epsilon_0) + \frac{dJ}{d\eps}\Bigr|_{\eps_0} \bigl(c_r V_0 (a+\ad) + \eps_d \cos{\w_d t}\bigr)  
  \\ + \frac{1}{2} \frac{d^2J}{d\eps^2}\Bigr|_{\eps_0} \bigl(c_r V_0 (a+\ad) + \eps_d \cos{\w_d t}\bigr)^2 + \ldots
 \end{multline}

The Hamiltonian for a qubit-resonator system is $H_{QR}=\hbar \w_r \ad a + \tfrac{1}{2}J(\eps) \sigma_z $. We move to an interaction picture with respect to $ H_0 =\hbar \w_d \ad a + \tfrac{1}{2} \tilde{J}(\eps_0)\sigma_z $, where $\tilde{J}=J(\eps_0)+\frac{1}{2} \left.\frac{d^2 J}{d \epsilon^2}\right|_{\epsilon_1} \bigl(c_r^2 V_0^2 + \epsilon_d^2 /2)$ includes second-order corrections to the DC value of $J$. Considering the first two orders of expansion and averaging over oscillating terms yields: 
\begin{align} \label{eq:H}
H_{\rm int} &=\hbar \Delta \ad a + \frac{1}{4} \frac{d^2J}{d\eps^2}\Bigr|_{\eps_0} \bigl(c_r V_0 \epsilon_d (a+\ad) + 2 c_r^2 V_0^2 \ad a \bigr) \sigma_z 
\nonumber \\
&=\hbar \Delta \ad a + \tfrac{1}{2}g(a+\ad)\sigma_z  + \tfrac{1}{2}\chi \ad a \sigma_z, 
\end{align}
where $ \Delta = \w_r - \w_d$ is the detuning, and $g =\frac{1}{2} \frac{d^2J}{d\eps^2}\bigr|_{\eps_0} c_r V_0 \epsilon_d$ and $\chi =  \frac{d^2J}{d\eps^2}\bigr|_{\eps_0} c_r^2 V_0^2 $ are coupling strengths.  The second coupling $\chi$ is smaller than $g$ by a factor of $c_r V_0 / \epsilon_d \ll 1$ for the optimal drive that we will consider, and so we will ignore it for the remainder. Higher order terms do not add additional terms to the equation, they only change their relative magnitude. 

To create a two-qubit coupling, we now add a second qubit to the model at the opposite antinode of the resonator, and drive it at the same frequency $\w_d$ and $180 ^{\circ}$ out of phase as the first qubit, giving the two-qubit Hamiltonian $ 
H_{2} =\hbar \Delta \ad a + \frac{g_1}{2} (a+\ad) \sza + \frac{g_2}{2} (a+\ad) \szb $. The technique of driving two qubits two qubits in resonance to enlarge the longitudinal coupling to the resonator and in turn one another is also employed in \onlinecite{royer2017}. Following Roos et al.\cite{roos2008}, it can be shown that $H_{2}$
generates a time-dependent phase space displacement $ U(t) = \exp[-i\Delta \cdot t a^\dagger a] \hat{D}[\alpha(t)]\exp[\Phi_{12}\sza \szb]$, where $\hat{D}$ is a qubit-dependent displacement operator, $\alpha(t) =  (1-e^{i \Delta \cdot t})(g_1 \sza + g_2 \szb)/(2\hbar \Delta)$, and $\Phi_{12}(t)=\frac{g_1 g_2}{2\hbar^2 \Delta^2}(\Delta \cdot t - \sin \Delta \cdot t)$. When $\alpha(t)=0$, the resonator disentangles from the qubits. A CPHASE gate occurs on the qubits when $\Phi_{12}=\pi/4$. Together, this requires that $ \Delta  \cdot t_g = 2 \pi n$ and $ \frac{g_1 g_2}{2\hbar^2 \Delta}t_g = \pi/4 $, where $n$ is a positive integer. This yields a requirement on the detuning $\hbar \Delta = 2 \sqrt{n g_1 g_2}$, and a CPHASE gate time $t_g = \pi\hbar\sqrt{n/(g_1 g_2)}$. While $n=1$, corresponding to a single oscillation of the resonator, yields the fastest gate, we will also consider $n>1$ gates to allow compatibility with dynamical decoupling, described below.

We now turn to an analysis of the main decoherence processes of this gate.  There are two main sources of loss in the system: dephasing of the qubits and loss of photons from the resonator. A master equation that governs the time evolution of the total system is: 
\begin{equation} \label{rho}
\dot{\rho} = -i[H_{2},\rho] + 2 \kappa \mathcal{D}[a]\rho + \gamma_{\phi,1} \mathcal{D}[\sza]\rho/2+\gamma_{\phi,2} \mathcal{D}[\szb]\rho/2, 
\end{equation}
where $ \kappa=\w_r/(2Q)$ is the cavity decay rate, $\gamma_{\phi,i}=1/T_{2,i}$ is the dephasing rate of qubit $i$, and $\mathcal{D}[c]\rho=c \rho c^{\dagger}-c^{\dagger} c \rho /2 - \rho c^{\dagger} c /2$ is the usual damping superoperator.  We neglect $T_1$ effects because $T_1$ exceeds $T_2$ by several orders of magnitude in most $S$-$T_0$ systems. We note that, while $T_2$ is limited by charge noise with a $1/f$ spectrum, the damping superoperator implements white noise. As a result, we expect the fidelities from solving this master equation to be slightly lower than in an exact simulation based on $1/f$ noise.

In this master equation, it is straightforward to analytically solve the dephasing of the qubits because all terms in the Hamiltonian commute with $\sigma_z$.  Each qubit therefore dephases at the rate $\gamma_{\phi,i}$ throughout the gate.  For dephasing due to loss of photons from the resonator, it is illustrative to solve the master equation analytically using a quantum trajectory approach.   We make the simplifying assumption that $g_1 =g_2 = g$ and $\gamma_{\phi,1}=\gamma_{\phi,2}=\gamma_{\phi}$. In general, if the qubits differ, we can replace $g$ and $\gamma_{\phi}$ in equations below with the geometric mean of the terms in the different qubits. The resulting dephasing process on the two qubits can be viewed as a stochastic process depending on whether an even or odd number of photons are lost from the cavity.  Odd numbers of lost photons result in a random $\pi$ phase shift of one of the qubits, whereas even numbers of lost photons result in a correlated $\pi$ phase shift of both qubits. The full analytical derivation of the system's evolution is provided in Appendix \ref{geomGate}. 

\section{Average Gate Fidelity} 
We can use the analytical derivation of the density matrix at the conclusion of the gate to find the average gate fidelity.  The average gate fidelity $\bar{F}_g$, as defined in Refs. \onlinecite{poyatos1997,nielsen2002}, is used as a measure to assess the performance of a noisy quantum gate compared with an ideal (unitary) gate.  It can be related to the fidelity of entanglement $F_e$, which is a simpler quantity to directly calculate, through the relation $\bar{F}_g = \frac{dF_e + 1}{d+1}$, where $d$ is the dimension of the Hilbert space ($d=4$ in the case of a two-qubit gate).  

We define a map to represent the action of our noisy CPHASE gate on a density matrix of two qubits as $\rho' = \mathcal{N}_{g}[\rho]$.  This map can be expressed as a matrix that acts on the space of density operators for 2 qubits.   To calculate the fidelity of entanglement for this two-qubit gate, we consider a maximally entangled state of \emph{four} qubits, with two of the qubits acted on by the gate.  Let $|\Psi\rangle=\tfrac{1}{2}\sum_{i,j=0,1}\ket{ij,ij}$ be the maximally entangled state of four qubits, with density matrix given by $\rho_{\Psi}=|\Psi\rangle\langle \Psi|=\tfrac{1}{4}\sum_{i,j,k,l=0,1}|ij,ij\rangle\langle kl,kl|$.  Then $\rho'_{\Psi} = (\mathcal{N}_{g}\otimes{I})[\rho_{_\Psi}]$ is a 4-qubit density matrix where the map $\mathcal{N}_{g}$ has been applied to qubits 1 and 2, and nothing has been done to qubits 3 and 4.  We define $\ket{\Psi'} = (U_{g} \otimes I) |\Psi\rangle$ as the state of the system after it has evolved under the ideal gate, so the fidelity of entanglement is 
\begin{equation}
F_e=\langle \Psi'|(\mathcal{N}_{t_g}\otimes{I})[\rho_{_\Psi}]|\Psi'\rangle = \langle \Psi|(\mathcal{N}'_{t_g}\otimes{I})[\rho_{_\Psi}]|\Psi\rangle \,,	
\label{eq:Fe1}
\end{equation}
where we have defined $\mathcal{N}'_{t_g} = \mathcal{U}_g^{-1} \circ \mathcal{N}_{t_g}$ with $\mathcal{U}_g^{-1}$ being the inverse ideal gate.  That is, $\mathcal{N}'_{t_g}$ describes only the noise in the gate.

This expression for the fidelity of entanglement can be made more explicit by using a trace-orthonormal basis of 2-qubit operators, such as the 2-qubit Pauli operators, to resolve the inner product of Eq.~(\ref{eq:Fe1}).  Let $\{\rho_\mu, \mu=1,\ldots,16\}$ be such a basis.  Then
\begin{equation}
F_e= \frac{1}{16}\sum_\mu {\rm Tr}\bigl[ \rho_\mu^\dag  \bigl(\mathcal{N}'_{t_g}(\rho_\mu)\bigr)\bigr] \,,
\label{eq:Fe2}
\end{equation}
We can then analytically calculate the gate fidelity using the basis of 2-qubit operators $\rho_\mu$ as initial states $\rho(0)$ in the solution of Eq.~(\ref{eq:soln3}).  Note that, for our numerical simulations, we use physical density matrices $\rho_k$ constructed from all two-qubit combinations of single qubit states $\ket{0},\, \ket{1},\, \frac{1}{\sqrt{2}}(\ket{0}+\ket{1}),\, \frac{1}{\sqrt{2}}(\ket{0}+i \ket{1})$ states, which can easily be shown to form an orthonormal basis of the operator space.

Given the analytical expression for the noisy CPHASE gate given in Eq.~(\ref{eq:soln3}), and including qubit dephasing as described by Eq.~(\ref{eq:deph}), we can analytically calculate the average gate fidelity.  Specifically, the terms in Eq.~(\ref{eq:soln3}) corresponding to zero, odd, and even numbers of lost photons provide Kraus operators for an operator product expansion of the map $\mathcal{N}_{t_g}$.  The average gate fidelity evaluates to  
\begin{equation}
\bar{F_g} = \frac{1}{10}(4 +4 b(t_g) e^{-\gam t_g}+(b(t_g)^4 + 1)e^{-2 \gam t_g})\,,
\end{equation}
where $b(t_g)$ represents the effect of photons lost from the resonator during the gate and remaining in the resonator at its completion and is defined in \eqref{eq:b} If we take the first order Taylor expansion of this around $\kappa t_g$ and $\gam t_g$, we find: 
\begin{equation}
1 - \bar{F_g} \approx \tfrac{4}{5}(\gam t_g + \kappa t_g/(2 n)).
\end{equation}

In the limit of small dephasing, the optimal gate time is $t_g=\sqrt n \pi \hbar/g$, and the corresponding gate fidelity $\bar{F_g}$ becomes
\begin{equation}
  1 - \bar{F_g} \approx \frac{4\sqrt{n} \pi}{5 g}(\gam + \kappa/(2 n)) = \frac{8 \sqrt{n} \pi \eps_0^2}{5 c_r J \sqrt{\hbar Z_r} \w_r \eps_d }(\gam + \frac{\omega_r}{4 n Q})\,.
\end{equation}

This simple expression for the fidelity enables us to find the optimal values of $J$ and $\eps_d$, giving a clearer picture of how the fidelity of the gate depends on the resonator's parameters and charge noise in the system.

\section{Effect of Charge Noise on Optimal Drive and Fidelity}

To optimize the parameters for driving the qubits with respect to their noise baths, we consider how spin qubits dephase under the influence of charge noise \cite{dial2013,cywinski2008}. Charge noise has a power spectrum described by $S(f)=S_{\eps}/f^{\beta}$, where $\beta$ is between 0.6 and 1.4 for a ``1/$f$'' spectrum\cite{kogan2008}. In the singlet-triplet qubit, we find that the noise is best understood as fluctuating charges affecting the chemical potentials of the quantum dots, which can be quantified by $S_{\eps}$, a function of the substrate, the dots' geometry and the qubit's wave function. To understand how this couples to the qubits' splitting, we perform a Taylor expansion similar to that in \ref{eq:J}, but we include an error term, $\delta \eps,$ that is time-dependent but negligible at the resonator's frequency due to the 1/$f$ nature of charge noise. Retaining terms linear in $\delta \eps$ in the third-order expansion and setting terms oscillating at the resonator frequency, we find: 
\begin{equation} 
\delta J \approx \delta \eps \Bigl(\left.\frac{dJ}{d\eps}\right|_{\eps_0} + \frac{1}{4} \left.\frac{d^3J}{d\eps^3}\right|_{\eps_0} \epsilon_d^2 \Bigr).
\label{eq:Jstat}
\end{equation} 
We can then find $S_J$ using the relation $S_J=1/2 (\frac{\delta J}{\delta \eps})^2 S_{\eps}$.  Empirical studies show that $J$ is an exponential of $\eps$, $J(\eps)\approx J_0 \exp(\eps/\eps_a),$ where $J_0$ and $\eps_a$ are tuning-dependent but can be treated as constants throughout multiple experiments where tuning is not substantially changed. 

We consider the possibility of employing dynamical decoupling pulses, despite the added complexity and requirement to perform the gate during multiple resonator oscillations, as they dramatically improve coherence times. Most gains come from a single echo, applied half-way through the gate time, which increases coherence times by about a factor of 30 \cite{dial2013}. For this reason, previous implementations of two-qubit gates in this system \cite{shulman2012,nichol2017} have employed a Hahn echo or rotary echo. Both gates rely on the same $\sigma_z \otimes \sigma_z$ interaction, so by performing simultaneous echoes on the qubits, we cancel noise but not the coupling. We intend to use the same technique in performing this gate, but to do this without affecting the CPHASE gate requires more care when the qubits are entangling with a resonator. One must perform echo $\pi$ pulses when the qubit is fully disentangled with the resonator, which can be achieved by performing the gate over multiple oscillations of the resonator.

To understand how $T_2$ varies with $J$, we consider previous work studying the effect of dynamical decoupling for colored noise \cite{cywinski2008}. If we implement dynamical decoupling with Gaussian charge noise, the resultant decay of qubit coherence takes the form $A(t) \propto \exp(-(t/T_2)^{1+\beta})$. In general, $T_2 = m^{\beta}/(\eta S_{\eps}^{1/(\beta+1)})$, where $\eta$ is a constant of order 1 and $m$ is the number of pulses performed. For certain pulse types, $\eta$ can be solved for analytically; for instance, for a Hahn echo, $\eta = \frac{1}{(2 \pi)}(2^{1-\beta}-1)\Gamma(-1-\beta)\sin(\frac{\pi \beta}{2})$. Combining the equations above, we find

 \begin{equation} 
 1/T_2 = \left(\eta S_{\eps} \frac{J^2}{\eps_a^2}\left(1 + \frac{\eps_d^2}{4 \eps_a^2}\right)^2 \right)^\frac{1}{\beta+1} \equiv \gamma_{\phi,0}\left(1 + \frac{\eps_d^2}{4 \eps_a^2}\right) ^\frac{2}{\beta+1},
 \end{equation}
 where we've defined the term $\gamma_{\phi,0}$ to represent the dephasing rate at a given value of $J$ with no drive applied. For the values of $S_{\eps}$ and $\beta$ used in simulations for this paper, and with no drive applied, $T_2(J)$ takes on the same set of values as in Ref \onlinecite{dial2013}. We optimize $\eps_d$ assuming exact $1/f$ noise to simplify the solution and from numerical solutions we find that this gives an accurate result, with the infidelity about 2\% lower when the exact form is used. 
 
 Using the model of the spectrum for charge noise with $\beta<1$, we can find optimal values $\eps_{d,{\rm opt}}$ and $J_{\rm opt}$ for the drive : $\eps_{d,\rm opt}= 2 \eps_a \sqrt{1+\frac{\kappa}{2 n \gamma_{\phi,0}}}$ and $J_{\rm opt} = (\frac{\beta}{2 n(1-\beta)} \kappa)^{(1+\beta)/2} \frac{\eps_a}{\sqrt{S_{\eps} \eta}}.$ The value $J_{\rm opt}$ has a strong dependence on $\beta$, going to $\infty$ for true $1/f$ noise and 0 for white noise. Because $J$ is limited to between about 50 MHz, where it becomes too small to drive effectively, and tens of GHz, we will not always be able to achieve the optimal value, but for values of $\beta\approx0.7$ that have been measured previously, those limits are not reached \cite{dial2013}.  Upon substituting in the optimal values $\eps_{d,{\rm opt}}$ and $J_{\rm opt}$, we find that 
 \begin{equation}
 \label{eq:InfidelityEst}
 1 - \bar{F_g} \approx \frac{4/5 \pi \left(\frac{2^3 n}{\beta}\right)^{\beta/2}(1-\beta)^{(1-\beta)/2} \sqrt{S_{\eps} \eta}}{\sqrt{\hbar Z_r}c_r Q^{(1-\beta)/2} \w_r^{(1+\beta)/2}}.
 \end{equation}


We note that (\ref{eq:InfidelityEst}) does not apply for $n=1$, because there $S_{\eps}$ takes on a much larger, low-frequency value.

\section{Expected Gate Performance}
To estimate the gate time and fidelity, we now look at the range of possible values taken by the parameters in the above expressions.  We first consider the impedance of the resonator.  While transmission lines relying on magnetic inductance are limited to approximately the impedance of free space, $Z \approx 377 \, \Omega$, kinetic inductance has no such physical limitation. Kinetic inductance, which arises from the inertia of electrons, can be found in several types of superconducting devices, including nanowires formed from type II superconductors and chains of SQUIDs. Using superconducting nanowires has yielded impedances up to 4000 $\Omega$ with quality 200,000 and chains of SQUIDs up to 50,000 $\Omega$ \cite{samkharadze2016,bell2012a}. Such large impedances preclude addressing or measuring the qubit using the resonator. The qubits we consider, however, allow for universal control and 98\% measurement fidelity independent of the resonator, and this has the added benefit of not requiring additional high frequency lines for resonators. High-impedance nanowires are typically much more compact than traditional resonators: for instance, a 5 GHz nanowire is of order 1 mm in length, but because it is only 100 nm wide, it can be folded so that it occupies a $(20 \mu \mathrm{m})^2$ area, or it can be extended to transport information over longer distances. Because $S$-$T_0$ qubits are typically a few $\mu \mathrm{m}^2$ in size, this retains the small size of a quantum dot based quantum processor, necessary for scaling to large numbers on chip. 

\begin{figure}
	\includegraphics[scale=0.28]{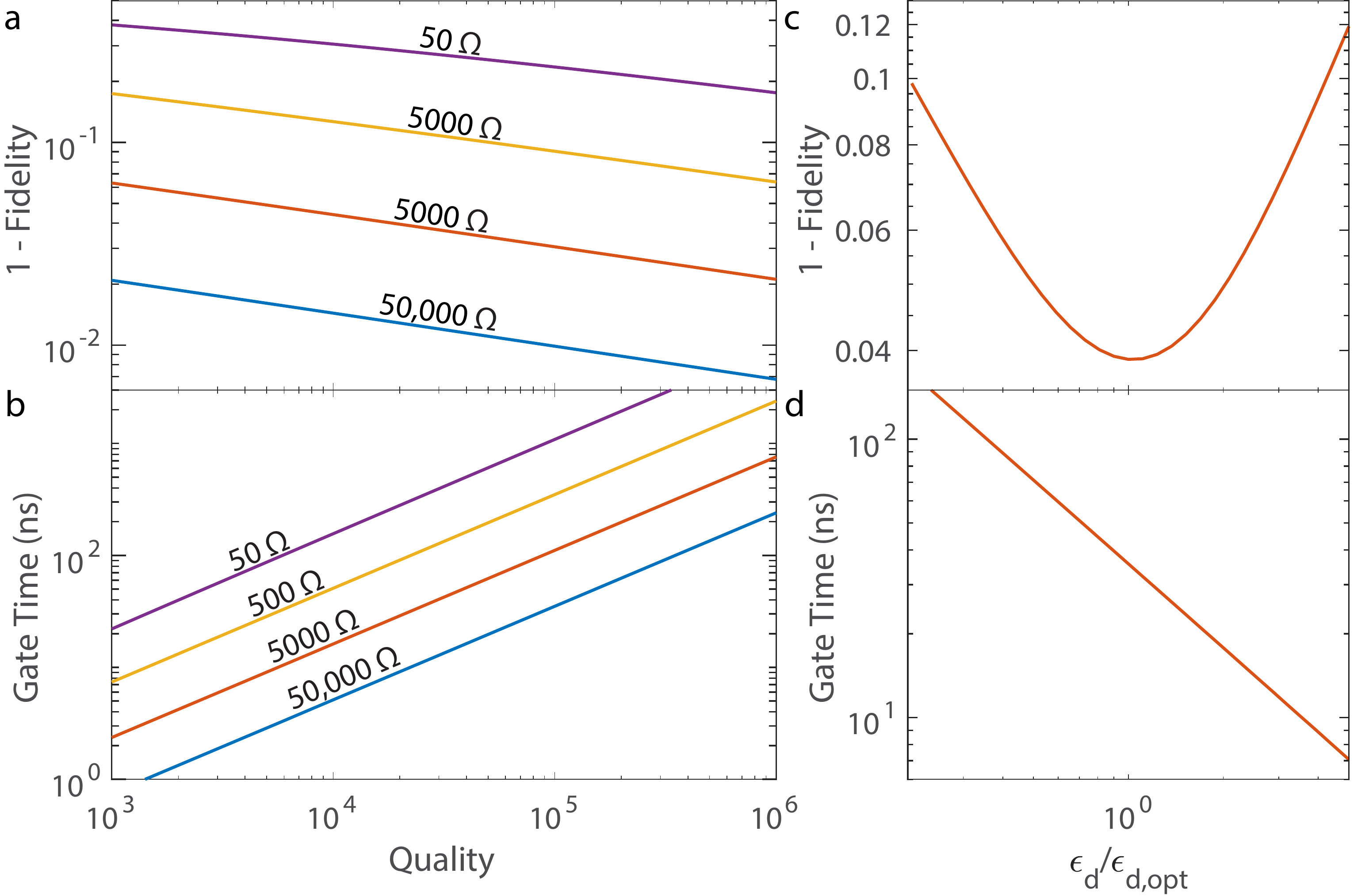}
	\caption{\textbf{a}, Average gate fidelity simulated by a numeric solution of the master equation with optimal values of J and $\eps_d$ as a function of resonator quality for resonator impedances of 50 $\Omega$ (standard), 500 $\Omega$ (maximum for magnetic impedance), 5000 $\Omega$ (typical nanowire), 50,000 $\Omega$ (SQUID array). \textbf{b}, CPHASE gate time for the same parameters as in a. \textbf{c} and \textbf{d}, Average gate fidelity and gate time as $\eps_d$ is varied around its optimal value with $Z_r = 5000 \; \Omega$ and $Q=20,000$. As quality increases, the gate time for the maximum fidelity gate increases as well, but by adjusting the drive, faster gate times can be achieved with minor loss of fidelity.  
		\label{fid2d}}
\end{figure}      

We next turn to the qubit-resonator coupling. Because the $S$-$T_0$ qubit has a dipole-like charge distribution, to generate a large coupling, we must maximize the coupling to one of the dots comprising the  qubit while minimizing the coupling to the other dot. This is quantified by the lever arm, $c_r = C_{R,r} / C_{R} - C_{L,r} / C_{L}$, with $C_{i,r}$ the capacitance between the resonator and each dot and $C_{i}$ the total capacitance of each dot. COMSOL simulations predict its value can range from 0 to 0.3, with $c_r = \, 0.05$ for previously measured functioning $S$-$T_0$ qubits in GaAs. In \figref{device}{b-c}, we show two proposals for resonator and qubit geometries in silicon-silicon germanium (Si-SiGe) quantum well devices.  In \figref{device}{c}, the resonator is at the same layer as the depletion gates and laterally adjacent to the near quantum dot, for which we predict $c_r$ to be 0.25. This large coupling relies on precise knowledge of the location of the dot, as a small shift would substantially reduce the coupling. In \figref{device}{b}, we replace one of the accumulation mode gates, which are vertically separated by an additional oxide layer, with the resonator, for which we predict $c_r$ to be 0.02. An ideal coupler would be as large as the gate in \figref{device}{b}, making it robust to different quantum dot locations, but would need to be separated by a much thinner layer of oxide, as has been used in recent experiments with aluminum gates separated with alumina \cite{mi2017}. 
        
We perform simulations of the density matrix to predict the expected average gate fidelity and gate time, as well as to confirm the power law behavior in Eq.~\eqref{eq:InfidelityEst}. To solve the differential equation for the density matrix of the system, we use a 4th order Runge-Kutta equation. The resonator can be populated up to 5 photons; the 5th level has a population of lower than $10^{-4}$ for all simulations. We use time spacing of 20 ps unless this gives more than 10,000 or under 200 time steps, in which case we use those limiting values. We use parameters $S_{\eps} = 1.4 \times 10^{-16}\, \mathrm{eV}^2/\mathrm{Hz}^{1-\beta}$, $\beta = 0.67$, $c_r=0.18$, the lever arm for devices that have been measured in GaAs, $w_r/(2\pi)= 6.5$ GHz and $\eta=0.086$, which corresponds to a Hahn echo pulse. The value of $S_{\eps}$ was measured in a GaAs device, as it has not available for any silicon devices at this time. 

In \figref{fid2d}{a-b}, we plot the simulated fidelity as a function of resonator quality and impedance. Fidelity ranges from 96\% to 99.3\%, an improvement of a factor of 5 to 25 compared to the maximum Bell state fidelity of $72\%$ achieved for the static capacitive gate reported in Ref.~\onlinecite{shulman2012}. The dephasing experienced by the qubit increases by a factor of close to 3; dephasing through the resonator increases the total rate by only about 50\%, however driving $J$ doubles the total noise. Gate times near 10 ns are readily achievable, as can be seen by inspection of \figref{fid2d}{b,d}. The dependence of $\eps_{d,{\rm opt}}$ and $J_{\rm opt}$ on the resonator's quality and the qubits' noise bath acts to equalize the effective noise from the two noise sources, the qubits and the resonator, for whatever the absolute values of noise may be. For instance, an increase in $S_{\eps}$ requires that we decrease $J$ and $\eps_d$ to keep the total noise originating with the qubit constant. As a result, the optimal gate time increases with quality, as that dictates a reduction of $J_{\rm opt}$ and $\eps_{d,{\rm opt}}$. We can decrease the gate time at the cost of fidelity, as is shown in \figref{fid2d}{c-d}. Here, we sweep $\eps_d$ while keeping all other parameters constant, and see that gate time has an inverse linear dependence on it, while the infidelity has a quadratic relationship on it, so the gate time can be substantially decreased without a corresponding excessive decrease in the fidelity. We note that the values of $g$ we find are generally larger than those in recent experimental works \cite{mi2018,landig2017,samkharadze2018}, which we attribute to increased coupling to charge in the singlet-triplet qubit and the ability to increase $g$ by driving the qubit in our model. 

While our focus has been on the coupling between two qubits, it is straightforward to include more qubits. Because the resonator does not require additional wiring, incorporating a resonator for each qubit pair does not impose scaling challenges. Another benefit of using resonators for two-qubit gates arises from the relative ease of fabricating and characterizing resonators compared to spin qubits. Improving the static capacitive gate's fidelity requires reducing charge noise in the system, which remains poorly understood and is challenging and time-consuming to measure. By comparison, resonator fabrication is an area of extremely active research, and dozens of resonators can be made and tested at once. High-impedance resonators will enable entangling gates to be performed in noisier samples than the pristine GaAs heterostructures that have been used in the past, an asset in the endeavor to scale to larger numbers of qubits.

We acknowledge useful discussions with and feedback from Bert Halperin, Michael Shulman, John Nichol and Di Wei. This work was supported by the Army Research Office grants W911NF-15-1-0203 and W911NF-17-1-024, and by the Australian Research Council project numbers CE110001013 and CE170100009.

\appendix
\section{Lever Arms}
\setcounter{equation}{0}

For singlet-triplet qubits, it is common to define $\eps=\frac{V_L-V_R}{2}$, with $V_L$ and $V_R$ the voltages on the RF gates directly over the left and right quantum dots, because that can be directly set. Here, we have defined $\eps=\mu_L-\mu_R$, the difference in chemical potential between the left and right dots. Two terms are affected by this change of definition: $\frac{dJ}{d\eps}$ and $S_{\eps}$. To convert these values, we first define $V_{\eps} = \frac{V_L-V_R}{2}$. Then $\frac{dJ}{d\eps}=\frac{dJ}{dV_{\eps}}\frac{dV_{\eps}}{d\eps}$ and $S_{\eps}=S_{V_{\eps}} (\frac{d\eps}{dV_{\eps}})^2$. We define $C_i, \, i=\{L,R\}$, to be the total capacitance of the left and right quantum dot, and $C_{i,j},\,i,j=\{L,R\}$ to be the capacitance between dot $i$ and gate $j$. Then $\frac{d\eps}{dV_{\eps}}=e \left( \frac{C_{LL} - C_{LR}}{C_L} - \frac{C_{RL} - C_{RR}}{C_R}\right) $, where $e$ is an electron charge. To find these capacitances, we perform a COMSOL simulation using a CAD file of the device, which allows us to extract a capacitance matrix for the system. We find that for the device in \cite{dial2013}, with which $S_{\eps}$ was measured, $\frac{d\eps}{dV_{\eps}}=0.182e$, which allows us to transform the power spectrum given in the paper, in terms of voltages on the gates, to the power spectrum of the dots' chemical potentials.

\section{Geometric Phase Gate} 
\label{geomGate}
\subsection{Ideal gate operation}

The ideal operation of our phase gate has been widely discussed in the ion trapping literature~\cite{leibfried2004,roos2008}. We now give a brief outline of these results so that the discussion here is self-contained and to establish notation. This discussion follows~\onlinecite{roos2008} closely.

Consider first a driven oscillator of oscillation frequency $\omega_r$. For our purposes it will be convenient to work in an interaction picture at the oscillator frequency.  In this picture, there will only be a linear drive of the oscillator, which makes it easier to find the explicit time-evolution operator.  A linear drive of the oscillator takes the general form
\begin{equation}
H_{\rm int}=  -i[E^*(t)a-E(t)a^\dagger] , \label{eq:directdrive}
\end{equation}
where $E(t)$ is a complex amplitude describing the linear drive.

The resulting time-evolution operator generated by this Hamiltonian satisfies the differential equation
\begin{equation}
\frac{d}{dt} U (t)=-iH_{\rm int}(t) U(t)/\hbar.\label{eq:displacementdynamics}
\end{equation}
There is a particularly straightforward solution to this equation using Glauber's displacement operators, $D[\alpha]=\exp\left(\alpha a^\dagger -\alpha ^* a\right)$.
The time-evolution operator for $H_{\rm{int}}$ is
\begin{equation}
U(t)=D[\alpha(t) ]\exp[\Phi(t)],
\end{equation}
where
\begin{align}
&\alpha(t)=\int_0^tdt' E(t')/\hbar 
\\
 &\Phi(t) ={\rm Im} \int_0^tdt' E(t') \int_0^{t'}dt'' E^*(t'')/\hbar^2.
\end{align}
This can be checked using standard facts about displacement operators. The phase factor $\Phi$ can be interpreted as the area enclosed in phase space by the oscillator dynamics; this can be seen by modifying the arguments at the end of the supplementary material of \onlinecite{leung2018} and is discussed in \onlinecite{leibfried2004}. 

We are interested in a system of two qubits longitudinally coupled to a single oscillator, in which case the interaction picture Hamiltonian involves only qubit-state-dependent forces on the oscillator and can be written
\begin{equation}
H_{\rm int}=  -i[E^*_1(t)\sigma_{z1}+E^*_2(t)\sigma_{z2}]a + {\rm h.c.} \label{eq:2qubitdrive}
\end{equation}
As noted in \onlinecite{roos2008}, because $\sigma_{zi}$ is a constant of the motion and commutes with $a$ and $a^\dagger$, we can obtain the solution for the full unitary by replacing the drive amplitude $E(t)$ by the qubit operator $E_1(t)\sigma_{z1}+E_2(t)\sigma_{z2}$ everywhere in the solution. Realistically, there would also be some direct drive of the oscillator of the form given by equation (\ref{eq:directdrive}), but we defer consideration of this to a later section. The resulting time-evolution operator generated by this Hamiltonian is
\begin{equation}
U(t)=\hat{D}[\alpha_1(t)\sigma_{z1}+\alpha_2(t)\sigma_{z2} ]\exp[\Phi_{12}(t)\sigma_{z1}\sigma_{z2}]\exp[\Phi_{0}(t)],
\end{equation}
where
\begin{equation}
\alpha_j(t)=\int_0^tdt'E_j(t')/\hbar 
\end{equation}
\begin{align}
\nonumber
\Phi_0=&{\rm Im}
 \left(\int_0^tdt'E_1(t')\int_0^{t'}dt''E_1^*(t'') \right. 
 \\ 
 &+\left. \int_0^tdt'E_2(t')\int_0^{t'}E^*_2(t'')\right)/\hbar^2,
\\ \nonumber
\Phi_{12}=&{\rm Im} (\int_0^tdt'E_1(t')\int_0^{t'}dt''E_2^*(t'') 
\\ 
&+ \int_0^tdt'E_2(t')\int_0^{t'}E^*_1(t''))/\hbar^2.
\end{align}
We have introduced the notation
\begin{eqnarray}
\hat{D}[\alpha_1\sigma_{z1}+\alpha_2\sigma_{z2} ]=& \nonumber
\exp\left[(\alpha_1\sigma_{z1}+\alpha_2\sigma_{z2})a^\dagger \right. 
\\
 &\left. -(\alpha^*_1\sigma_{z1}+\alpha^*_2\sigma_{z2}) a\right] .
\end{eqnarray}

$\hat{D}$ is a qubit-state-dependent displacement of the oscillator, so it is an entangling operation that acts on both qubits as well as the oscillator. We reserve the notation $D[\alpha]$ for the usual displacement operators that act nontrivially on the oscillator alone.

In the main text, we worked in an interaction picture with respect to the drive frequency $\omega_d$, which has the Hamiltonian described by \eqref{eq:H}. When we move to the interaction picture used above, we find that the resulting interaction picture Hamiltonian is a special case of the one in (\ref{eq:2qubitdrive}):
\begin{equation}
H_{\rm int}=  -\frac{i}{2}\left[g_1 e^{-i\Delta t}\sigma_{z1}+g_2(t)e^{-i\Delta t}\sigma_{z2}\right]a + {\rm h.c.}
\end{equation}
Thus for the purposes of the main text we are interested in the case where $E_j=g_je^{i\Delta t}/2$.  In this case, 
\begin{eqnarray}
\alpha_j(t)&=& i\frac{g_j}{2\hbar\Delta} \left(1-e^{i\Delta t}\right)  \\
\Phi_{12} &=& \frac{g_1g_2}{2\hbar^2\Delta^2} \left(\Delta t -\sin \Delta t\right), 
\end{eqnarray}
and we want to choose a gate time such that $|\Delta | t_g=2\pi n $, for which $\alpha_j=0$ and the gate acts only on the qubits with $\Phi_{12}=\pm ng_1g_2/2\hbar^2\Delta^2$, as stated in the main text.

Ultimately, we are interested in the time evolution operator generated by the Hamiltonian $H_2$ in the original interaction picture. This is straightforward to find from the textbook discussion of the interaction picture:
\begin{align}
U(t)=&\exp[-i\Delta t a^\dagger a]\hat{D}[\alpha_1(t)\sigma_{z1}+\alpha_2(t)\sigma_{z2}] \nonumber
\\
&\exp[\Phi_{12}(t)\sigma_{z1}\sigma_{z2}]\exp[\Phi_{0}(t)].
\end{align}

\subsection{Dephasing due to cavity decay}

We now consider the effects of dephasing on the two-qubit gate. It is illustrative to first study the effect of the decay of photons from the resonator on its own and how this decay affects the geometric phase acquired. Qubit dephasing is straightforward to incorporate later because it commutes with the gate. The restricted master equation is:
\begin{equation}
\dot{\rho}=-i[H_{\rm int},\rho]+2\kappa\mathcal{D}[a]\rho\,. \label{eq:masterequation}
\end{equation}

We first write the solution of this master equation for the oscillator initially in its ground state and the pair of qubits in an initial state $\rho_q$, so that the initial state of the system is   $\rho(0)=|0\rangle\langle0|\otimes \rho_q$. At the conclusion of the gate, we average (trace) over the state of the cavity. Initial coherent and thermal states of the cavity are also tractable and will be discussed in a later section. We will restrict attention to the case where $E_1(t)=E_2(t)$; one could write analogous expressions with different driving on the two qubits, but certain simplifications would not occur. 

We begin by defining $\alpha(t)$ so that it satisfies
\begin{equation}
\dot{\alpha}=-\kappa\alpha+E(t)/\hbar\,, \label{eq:alphadefn}
\end{equation}
describing the amplitude of the oscillator subject to the decay rate $\kappa$.
Associated with this differential equation we define a phase, analogous to the two-qubit phase acquired in the lossless case,
\begin{equation}
\Phi_{12}(t)=2{\rm Im} \int_0^tdt' E(t')\alpha^*(t')/\hbar. \label{eq:phi0defn}
\end{equation}
Continuing the analogy, we define a two-qubit gate performed at time $t_g$,
\begin{equation}
U_g=\exp[i\Phi_{12}(t_g)\sigma_{z1}\sigma_{z2}]. \label{eq:ugdefn}
\end{equation}

It can be shown that state of the qubits after evolving according to the master equation (\ref{eq:masterequation}) and averaging over the oscillator is then
\begin{equation}
\rho_q(t_g)= U_g\mathcal{E}[\rho_q(0)]U_g^\dagger, \label{eq:soln}
\end{equation}
where $\mathcal{E}$ is a correlated dephasing process associated with residual excitations in the oscillator at the end of the gate and the excitations lost from the oscillator during the gate. The dephasing process $\mathcal{E}$ commutes with the action of the unitary gate $U_g$ so the order of operations in equation (\ref{eq:soln}) can be changed as required. 

The correlated dephasing process can be described as
\begin{equation}
\mathcal{E}[\rho_q(0)]=\rho_0+\rho_{\rm odd}+\rho_{\rm even}. \label{eq:soln3}
\end{equation}
The three contributions correspond to, respectively, zero photons, an odd number of photons, and an even number of photons either lost from the cavity or remaining in it at the end of the pulse. They are defined as follows:
\begin{align}
&\rho_0 = [(1+b)I - (1-b)\sza \szb] \nonumber
\\
& \qquad \, \rho_q(0)[(1+b)I-(1-b)\sza\szb]/4 \,, \label{eq:Kraus1}\\
&\rho_{\rm odd} = \frac{1}{2}\left( 1 - b^4 \right) [\sza + \szb]\rho_q(0)[\sza + \szb]/4\,, \label{eq:Kraus2}\\
& \rho_{\rm even} = \frac{1}{2}\left( 1 - b^2 \right) ^2 [I + \sza \szb]\rho_q(0)[I+\sza\szb]/4\,,\label{eq:Kraus3}
\end{align}
where 
\begin{equation}
b(t_g) = \exp\left(-4 \kappa \int_0^{t_g} dt' |\alpha(t')|^2 - 2 |\alpha(t_g)|^2\right) \label{eq:b}.
\end{equation}
Together, the equations (\ref{eq:Kraus1}-\ref{eq:Kraus3}) provide a Kraus operator expansion for $\mathcal{E}$.

As in the ideal case, we are most interested in the specific drive where $E=ge^{i\Delta t}/2$. In this case we find
\begin{equation}
\alpha(t)=-\frac{g}{2\hbar}\frac{1}{i\Delta +\kappa}(e^{-\kappa t}-e^{i\Delta t}).
\end{equation}
Substituting this into \eqref{eq:b} yields
\begin{align}
b(t_g)= & \exp \left[ \frac{- \kappa t_g g^2}{\hbar^2 (\Delta^2 + \kappa^2)} + 
 \frac{g^2(\Delta^2 - \kappa^2)}{\hbar^2(\Delta^2 + \kappa^2)^2} (\cos \Delta t_g e^{- \kappa t_g}-1) 
 \right. \nonumber
  \\ 
 +& \left. \frac{2 g^2 e^{-\kappa t_g}  \kappa  \Delta}{\hbar^2(\Delta^2 + \kappa^2)^2} \sin \Delta t_g 
 \right]. 
\end{align} 

At $t_g = \pi\sqrt{n} \hbar/g$ and $\hbar \Delta = 2\sqrt{n}g$, this simplifies to $b(t_g) = \exp (-\pi \hbar \kappa / (2 g \sqrt{n} ))$ when $\kappa\ll \Delta$.

\subsection{Explicit solution to the master equation}

In this subsection we will justify the expression for the state of the qubits at the conclusion of the gate (\ref{eq:soln}). Recall that this arises from solving the dynamics according to the master equation (\ref{eq:masterequation}) with the initial state $\rho(0)=|0\rangle\langle 0|\otimes\rho_q$ and then averaging over the state of the oscillator. The correlated dephasing of the qubits described by (\ref{eq:soln3}) arises from both the operation of the gate and the entanglement of the qubits with the oscillator at the conclusion of the gate. These two contributions can be seen in the two terms in the exponent on the right hand side of equation \eqref{eq:b}. 

Our approach in this section will be to first state the solution $\rho(t)$ to the master equation (\ref{eq:masterequation}) with the initial condition given above. We will then outline an argument that verifies that this proposed solution satisfies the master equation (although we originally found the solution by solving the master equation explicitly, the details are unnecessarily complicated to repeat here). In the second step of the calculation, we will average over the oscillator to obtain the reduced density matrix $\rho_q$ for the qubits alone, recovering the claimed solution (\ref{eq:soln3}). Readers willing to trust this solution may wish to skip the rest of this subsection. 

It will be simpler to write various intermediate expressions as members of a one-parameter family of correlated dephasing processes $\mathcal{E}_b$ that we define as follows:
\begin{equation}
\mathcal{E}_b[\rho_q(0)]=\rho_0+\rho_{\rm odd}+\rho_{\rm even}. 
\end{equation}
The terms on the right hand side are defined as
\begin{align}
&\rho_0 = [(1+b)I - (1-b)\sza \szb] \nonumber
\\
& \qquad \, \times \rho_q(0)[(1+b)I-(1-b)\sza\szb]/4 \,, \label{eq:Kraus1}\\
&\rho_{\rm odd} = \tfrac{1}{2}\left( 1 - b^4 \right) [\sza + \szb]\rho_q(0)[\sza + \szb]/4\,, \label{eq:Kraus2}\\
& \rho_{\rm even} = \tfrac{1}{2}\left( 1 - b^2 \right) ^2 [I + \sza \szb]\rho_q(0)[I+\sza\szb]/4\,,\label{eq:Kraus3}
\end{align}

For this to be a valid quantum operation, the parameter $b$ needs to satisfy $0\leq b\leq 1$. This definition has the useful property that
\begin{equation}
\mathcal{E}_{b_1}[\mathcal{E}_{b_2}[\rho]]=\mathcal{E}_{b_1b_2}[\rho]. \label{eq:semigroup}
\end{equation}
The value of $b$ for the solution is given in \eqref{eq:b}.

Given that notation, the solution to the master equation with our desired initial state is
\begin{equation}
\rho(t) = U_g \hat{D}[\alpha(\sigma_{z1}+\sigma_{z2}) ] \bigl(|0\rangle\langle 0|\otimes \mathcal{E}_{b_l} [\rho_q]\bigr)  \hat{D}^\dagger[\alpha(\sigma_{z1}+\sigma_{z2}) ] U^\dagger_g, \label{eq:soln0}
\end{equation}
where $\alpha$, and $U_g$ are defined above in equations (\ref{eq:alphadefn}), and (\ref{eq:ugdefn}) respectively and $b_l$, which represents the contribution to qubit dephasing arising from the \emph{loss} of excitations from the oscillator during the operation of the gate, is defined as
\begin{equation}
b_l(t)=\exp\left(-4\kappa \int_0^t dt' |\alpha(t')|^2\right).
\end{equation}

There are several ways to establish this solution. In our original approach we found analytical solutions to the quantum trajectory equations that describe the dynamics of the system as the oscillator emits excitations into the bath\cite{wiseman2010,carmichael1993} and then summed over the number of emissions and averaged over the various emission times. In this solution, $\rho_0$ is associated with trajectories in which no emissions occur, $\rho_{\rm odd}$ with an odd number of emissions, and $\rho_{\rm even}$ with an even number of emissions. While this calculation gives a nice physical picture, the details are tedious and to verify the solution we just need to check that it satisfies the master equation and has the appropriate initial condition as we do in the following. 

We begin by performing the well-known polaron transform, often used to analyze the master equations for longitudinal coupling of qubits to oscillators \cite{gambetta2008}. This transformation simplifies the dynamics by essentially decoupling the qubits and the oscillator. The ``polaron picture" density matrix is defined as follows:
\begin{equation}
\tilde{\rho}(t)= \hat{D}[-\alpha(\sigma_{z1}+\sigma_{z2}) ] \rho(t) \hat{D}^\dagger[-\alpha(\sigma_{z1}+\sigma_{z2}) ] .
\end{equation}
Since $\alpha(0)=0$, the initial condition for this master equation is $\tilde{\rho}=|0\rangle\langle 0| \otimes \rho_q(0)$.
Our proposed solution for $\tilde{\rho}(t)$, from (\ref{eq:soln0}), is
\begin{equation}
\tilde{\rho}(t) =  |0\rangle\langle 0|\otimes \left( U_g \mathcal{E}_{b_l} [\rho_q]U^\dagger_g\right). \label{eq:polaronsoln}
\end{equation}
In the polaron picture, the oscillator remains in the ground state at all times.

The master equation for $\tilde{\rho}$ is
\begin{widetext}
\begin{eqnarray}
\frac{d}{dt}\tilde{\rho} = \left(\frac{d}{dt}\hat{D}[-\alpha(\sigma_{z1}+\sigma_{z2})]\right)\rho \hat{D}^\dagger[-\alpha(\sigma_{z1}+\sigma_{z2})]  +\hat{D}[-\alpha(\sigma_{z1}+\sigma_{z2})]\rho  \left(\frac{d}{dt}\hat{D}[-\alpha(\sigma_{z1}+\sigma_{z2})]\right)^\dagger
+ \hat{D}[-\alpha(\sigma_{z1}+\sigma_{z2})]\left(\frac{d}{dt}\rho(t)\right) \hat{D}^\dagger[-\alpha(\sigma_{z1}+\sigma_{z2})].
\end{eqnarray}

 Standard techniques can be used to show that
\begin{align}
\frac{d}{dt}\hat{D}[-\alpha(\sigma_{z1}+\sigma_{z2})] = [(\dot{\alpha}^*a-\dot{\alpha}a^\dagger)(\sigma_{z1}+\sigma_{z2}) 
+(\dot{\alpha}\alpha^*-\dot{\alpha}^*\alpha)(\sigma_{z1}\sigma_{z2}+I)]\hat{D}[-\alpha(\sigma_{z1}+\sigma_{z2})].
\end{align}
\end{widetext}
After some calculations relying on the fact that the oscillator remains in its ground state, we find

\begin{eqnarray}
\dot{\tilde{\rho}} 
&=& -i[\tilde{H}_{\rm int},\tilde{\rho}]+2\kappa|\alpha(t)|^2\mathcal{D}[\sigma_{z1}+\sigma_{z2}]\tilde{\rho},\label{eq:polaron}
\end{eqnarray}

where
\begin{equation}
\tilde{H}_{\rm int}=  -i[E(t)\alpha^*(t)-E^*(t)\alpha(t)]\sigma_{z1}\sigma_{z2}  
\end{equation}
and $\alpha(t)$ is a solution to equation (\ref{eq:alphadefn}).

It is easy to check that $\tilde{\rho}(t)$ as given in (\ref{eq:polaronsoln}) satisfies the polaron picture master equation by substituting $\tilde{\rho}$ into the left and right hand side of (\ref{eq:polaron}) and checking that they match. This therefore shows that $\rho(t)$ is the correct solution of the original master equation.

The final step to find the state of the qubits $\rho_q(t)$ is to average over the oscillator by taking the partial trace 
\begin{align}
\nonumber
\rho_q(t) =& U_g \mathrm{Tr}_{\rm cav}\left[\hat{D}[\alpha(\sigma_{z1}+\sigma_{z2}) ] \right.
\\
& \times \left. \bigl(|0\rangle\langle 0|\otimes \rho'_q\bigr)  \hat{D}^\dagger[\alpha(\sigma_{z1}+\sigma_{z2}) ]\right] U^\dagger_g,
\end{align}
where $\rho'_q=\mathcal{E}_{b_l} [\rho_q]$. To facilitate the calculation of the partial trace note that
\begin{align}
\hat{D}[\alpha (\sigma_{z1}+\sigma_{z2} )] =& D[2\alpha ]\otimes |00\rangle\langle 00|+I \otimes  |01\rangle\langle 01| \nonumber
\\ 
&+I \otimes |10\rangle\langle 10|+D[-2\alpha]\otimes |11\rangle\langle 11| .
\end{align}
We can make use of the following identity for a pair of coherent states,
\begin{equation}
{\rm Tr}[|\alpha\rangle \langle \beta|] = \langle \beta|\alpha\rangle = e^{-|\alpha|^2/2-|\beta|^2/2+\beta^*\alpha}.
\end{equation}
It is then straightforward to verify that
\begin{align}
\nonumber
\mathrm{Tr}_{\rm cav} & \left[ \hat{D}[\alpha_1\sigma_{z1}+\alpha_2\sigma_{z2} ] \left(|0\rangle\langle 0|\otimes \rho'_q \right) \right.
\\
& \times \left. \hat{D}^\dagger[\alpha_1\sigma_{z1}+\alpha_2\sigma_{z2} ]\right] = \mathcal{E}_{b_e}[\rho'_q],
\end{align}
where we have defined
\begin{equation}
b_e(t)=\exp\left(-2\kappa |\alpha(t)|^2\right).
\end{equation}
$b_e$ is the contribution to the dephasing of the qubits arising from the \emph{entanglement} of the qubits with the oscillator at the conclusion of the gate.

Therefore, using (\ref{eq:semigroup}), we obtain the desired result
\begin{equation}
\rho_q(t_g) = U_g \mathcal{E}_b[\rho_q] U^\dagger_g,
\end{equation}
with $b(t)=b_l(t)b_e(t)$ given by equation \eqref{eq:b}. 

\subsection{Effect of direct oscillator drive and non-vacuum initial state}

It may seem overly restrictive to restrict the initial oscillator state to the ground state. In practice, we would like to understand the behavior of the gate for both initial coherent and initial thermal states of the oscillator. Likewise, in practice the longitudinal coupling of the qubits to the oscillator will involve some direct drive of the oscillator, and we would like to model this effect. In this section, we explain how to extend the solution to these cases. 

We consider an interaction picture Hamiltonian of the form 
\begin{equation}
H_{\rm int}=  -i[E^*_c(t)+E^*(t)\sigma_{z1}+E^*(t)\sigma_{z2}]a + {\rm h.c.} \label{eq:2qubitcavdrive}
\end{equation}
which includes a direct cavity drive $E_c$. We consider initial states of the form $\rho(0)=|\beta\rangle\langle\beta|\otimes \rho_q$, where $|\beta\rangle$ is a coherent state. Finding the solution for an initial coherent state also allows us to model thermal states by averaging over a Gaussian probability distribution for $\beta$.

In the limiting case where the longitudinal coupling is negligible, it is straightforward to show that the solution is $\rho(t)= |\alpha_c(t)\rangle\langle\alpha_c(t)|\otimes \rho_q$ where $\alpha_c(t)$ satisfies
\begin{equation}
\dot{\alpha}_c=-\kappa \alpha_c+E_c/\hbar,
\end{equation}
and $\alpha_c(0)=\beta$.  Following the approach used in the previous section, we define
\begin{equation}
\tilde{\rho}(t)= D[-\alpha_c]\rho(t) D^\dagger[-\alpha_c],
\end{equation}
so that $\tilde{\rho}(t)=|0\rangle \langle 0|\otimes \rho_q$ at all times for negligible longitudinal coupling.

Working with a master equation for $\tilde{\rho}$, we have
\begin{align}
\nonumber
\dot{\tilde{\rho}} =& \dot{D}[-\alpha_c]\rho D^\dagger[-\alpha_c] +D[-\alpha_c]\rho \left(\dot{D}[-\alpha_c]\right)^\dagger 
\\
& + D[-\alpha_c]\dot{\rho}(t) D^\dagger[-\alpha_c].
\end{align}

We find
\begin{align}
\nonumber
\dot{\tilde{\rho}} =& -(\dot{\alpha}^*_c a+\dot{\alpha}_c a^\dagger)\tilde{\rho} - \tilde{\rho}(\dot{\alpha}_c a^\dagger+\dot{\alpha}^*_c a) -i[-iE^*_ca+iE_ca^\dagger,\tilde{\rho}]
\\
 &\quad-i[H_{\rm int},\tilde{\rho}]+2\kappa\mathcal{D}[a+\alpha_c]\tilde{\rho} \\
=& -i[\tilde{H}_{\rm int},\tilde{\rho}]+2\kappa\mathcal{D}[a]\tilde{\rho},
\end{align}
where
\begin{align}
\nonumber
\tilde{H}_{\rm int}=&  -i[E_c^*(t)a-E_c(t)a^\dagger](\sigma_{z1}+\sigma_{z2}) 
\\
  &-i[\alpha^*_c(t) E_c^*(t)-\alpha_c(t)E_c(t)](\sigma_{z1}+\sigma_{z2}) \label{eq:qubitoscdrive}.
\end{align}

We see that the coherent state in the oscillator can be handled by adding a classical drive that affects only the qubits. Since this drive term commutes with the terms that couple the qubits and the cavity, we can infer the solution to this master equation from the solution with no extra drive by adding an appropriate local unitary $U_{1q}$ at the end of the calculation. 

Again, we are interested in the reduced state of the qubits at the conclusion of the gate, which is given by $\rho_{q}(t_g)={\rm Tr}_{\rm cav}[\rho(t_g)]$.  Because the displacement is a unitary that acts only on the oscillator, it does not affect the partial trace, so $\rho_{q}(t_g)={\rm Tr}_{\rm cav}[\tilde{\rho}(t_g)]$ and we can work entirely in this displaced picture to calculate the quality of the gate. We find as before 
\begin{equation}
\rho_q(t_g) = U_{1q}U_g \mathcal{E}_b[\rho_q] U^\dagger_g U^\dagger_{1q},
\end{equation}
with $b(t)$ given by equation \eqref{eq:b}.

\subsection{Adding qubit dephasing}

The above analysis considered only the restricted master equation that describes dephasing due to cavity decay.  As we noted above, it is straightforward to include qubit dephasing because all of the dephasing terms commute with one another. That is, the effect of intrinsic dephasing on each qubit can be described as a noise map on the input density matrix that acts independently of, and commutes with, the noise resulting from cavity decay.

As discussed in the main text, qubit dephasing for two qubits is described by a contribution to the master equation $\dot{\rho} = \gamma_{\phi,1} \mathcal{D}[\sigma_{z1}]\rho+\gamma_{\phi,2} \mathcal{D}[\sigma_{z2}]\rho$, and in the absence of coupling to the oscillator ($g=0$), this has the following solution:

\begin{align} \label{eq:deph}
\nonumber 
\rho_q(t)=&\mathcal{E}_q[\rho_q(0)]=(1-p_1)(1-p_2)\rho_q(0) 
\\ &+p_1(1-p_2) \sigma_{z1} \rho_q(0) \sigma_{z1} +p_2(1-p_1) \sigma_{z2} \rho_q(0) \sigma_{z2}
\nonumber
\\ & + p_1 p_2 \sigma_{z1}\sigma_{z2} \rho_q(0) \sigma_{z1}\sigma_{z2},
\end{align}

where $p_j=\tfrac{1}{2}(1-e^{-\gamma_{\phi,j} t}).$ This intrinsic dephasing of the qubits is independent of the coupling to the cavity because all terms commute with $\sigma_{z1}$ and $\sigma_{z2}$, so the overall solution is
\begin{equation}
\rho_q(t_g)= U_g\mathcal{E}[\mathcal{E}_q[\rho_q(0)]]U_g^\dagger. \label{eq:fullsoln}
\end{equation}

The structure of this solution would be unchanged if the intrinsic qubit dephasing were non-Markovian and therefore not described by a master equation. The effect of the dephasing would still be described by some single qubit dephasing process of the same form as (\ref{eq:deph}) with some values of $p_1$ and $p_2$, so (\ref{eq:fullsoln}) holds.
\bibliography{resBib}

\begin{thebibliography}{37}%
\makeatletter
\providecommand \@ifxundefined [1]{%
 \@ifx{#1\undefined}
}%
\providecommand \@ifnum [1]{%
 \ifnum #1\expandafter \@firstoftwo
 \else \expandafter \@secondoftwo
 \fi
}%
\providecommand \@ifx [1]{%
 \ifx #1\expandafter \@firstoftwo
 \else \expandafter \@secondoftwo
 \fi
}%
\providecommand \natexlab [1]{#1}%
\providecommand \enquote  [1]{``#1''}%
\providecommand \bibnamefont  [1]{#1}%
\providecommand \bibfnamefont [1]{#1}%
\providecommand \citenamefont [1]{#1}%
\providecommand \href@noop [0]{\@secondoftwo}%
\providecommand \href [0]{\begingroup \@sanitize@url \@href}%
\providecommand \@href[1]{\@@startlink{#1}\@@href}%
\providecommand \@@href[1]{\endgroup#1\@@endlink}%
\providecommand \@sanitize@url [0]{\catcode `\\12\catcode `\$12\catcode
  `\&12\catcode `\#12\catcode `\^12\catcode `\_12\catcode `\%12\relax}%
\providecommand \@@startlink[1]{}%
\providecommand \@@endlink[0]{}%
\providecommand \url  [0]{\begingroup\@sanitize@url \@url }%
\providecommand \@url [1]{\endgroup\@href {#1}{\urlprefix }}%
\providecommand \urlprefix  [0]{URL }%
\providecommand \Eprint [0]{\href }%
\providecommand \doibase [0]{http://dx.doi.org/}%
\providecommand \selectlanguage [0]{\@gobble}%
\providecommand \bibinfo  [0]{\@secondoftwo}%
\providecommand \bibfield  [0]{\@secondoftwo}%
\providecommand \translation [1]{[#1]}%
\providecommand \BibitemOpen [0]{}%
\providecommand \bibitemStop [0]{}%
\providecommand \bibitemNoStop [0]{.\EOS\space}%
\providecommand \EOS [0]{\spacefactor3000\relax}%
\providecommand \BibitemShut  [1]{\csname bibitem#1\endcsname}%
\let\auto@bib@innerbib\@empty
\bibitem [{\citenamefont {Petta}\ \emph {et~al.}(2005)\citenamefont {Petta},
  \citenamefont {Johnson}, \citenamefont {Taylor}, \citenamefont {Laird},
  \citenamefont {Yacoby}, \citenamefont {Lukin}, \citenamefont {Marcus},
  \citenamefont {Hanson},\ and\ \citenamefont {Gossard}}]{petta2005a}%
  \BibitemOpen
  \bibfield  {author} {\bibinfo {author} {\bibfnamefont {J.~R.}\ \bibnamefont
  {Petta}}, \bibinfo {author} {\bibfnamefont {A.~C.}\ \bibnamefont {Johnson}},
  \bibinfo {author} {\bibfnamefont {J.~M.}\ \bibnamefont {Taylor}}, \bibinfo
  {author} {\bibfnamefont {E.~A.}\ \bibnamefont {Laird}}, \bibinfo {author}
  {\bibfnamefont {A.}~\bibnamefont {Yacoby}}, \bibinfo {author} {\bibfnamefont
  {M.~D.}\ \bibnamefont {Lukin}}, \bibinfo {author} {\bibfnamefont {C.~M.}\
  \bibnamefont {Marcus}}, \bibinfo {author} {\bibfnamefont {M.~P.}\
  \bibnamefont {Hanson}}, \ and\ \bibinfo {author} {\bibfnamefont {A.~C.}\
  \bibnamefont {Gossard}},\ }\href {\doibase 10.1126/science.1116955}
  {\bibfield  {journal} {\bibinfo  {journal} {Science}\ }\textbf {\bibinfo
  {volume} {309}},\ \bibinfo {pages} {2180} (\bibinfo {year}
  {2005})}\BibitemShut {NoStop}%
\bibitem [{\citenamefont {Wu}\ \emph {et~al.}(2014)\citenamefont {Wu},
  \citenamefont {Ward}, \citenamefont {Prance}, \citenamefont {Kim},
  \citenamefont {Gamble}, \citenamefont {Mohr}, \citenamefont {Shi},
  \citenamefont {Savage}, \citenamefont {Lagally}, \citenamefont {Friesen},
  \citenamefont {Coppersmith},\ and\ \citenamefont {Eriksson}}]{wu2014}%
  \BibitemOpen
  \bibfield  {author} {\bibinfo {author} {\bibfnamefont {X.}~\bibnamefont
  {Wu}}, \bibinfo {author} {\bibfnamefont {D.~R.}\ \bibnamefont {Ward}},
  \bibinfo {author} {\bibfnamefont {J.~R.}\ \bibnamefont {Prance}}, \bibinfo
  {author} {\bibfnamefont {D.}~\bibnamefont {Kim}}, \bibinfo {author}
  {\bibfnamefont {J.~K.}\ \bibnamefont {Gamble}}, \bibinfo {author}
  {\bibfnamefont {R.~T.}\ \bibnamefont {Mohr}}, \bibinfo {author}
  {\bibfnamefont {Z.}~\bibnamefont {Shi}}, \bibinfo {author} {\bibfnamefont
  {D.~E.}\ \bibnamefont {Savage}}, \bibinfo {author} {\bibfnamefont {M.~G.}\
  \bibnamefont {Lagally}}, \bibinfo {author} {\bibfnamefont {M.}~\bibnamefont
  {Friesen}}, \bibinfo {author} {\bibfnamefont {S.~N.}\ \bibnamefont
  {Coppersmith}}, \ and\ \bibinfo {author} {\bibfnamefont {M.~A.}\ \bibnamefont
  {Eriksson}},\ }\href {\doibase 10.1073/pnas.1412230111} {\bibfield  {journal}
  {\bibinfo  {journal} {Proceedings of the National Academy of Sciences}\
  }\textbf {\bibinfo {volume} {111}},\ \bibinfo {pages} {11938} (\bibinfo
  {year} {2014})}\BibitemShut {NoStop}%
\bibitem [{\citenamefont {Eng}\ \emph {et~al.}(2015)\citenamefont {Eng},
  \citenamefont {Ladd}, \citenamefont {Smith}, \citenamefont {Borselli},
  \citenamefont {Kiselev}, \citenamefont {Fong}, \citenamefont {Holabird},
  \citenamefont {Hazard}, \citenamefont {Huang}, \citenamefont {Deelman},
  \citenamefont {Milosavljevic}, \citenamefont {Schmitz}, \citenamefont {Ross},
  \citenamefont {Gyure},\ and\ \citenamefont {Hunter}}]{eng2015}%
  \BibitemOpen
  \bibfield  {author} {\bibinfo {author} {\bibfnamefont {K.}~\bibnamefont
  {Eng}}, \bibinfo {author} {\bibfnamefont {T.~D.}\ \bibnamefont {Ladd}},
  \bibinfo {author} {\bibfnamefont {A.}~\bibnamefont {Smith}}, \bibinfo
  {author} {\bibfnamefont {M.~G.}\ \bibnamefont {Borselli}}, \bibinfo {author}
  {\bibfnamefont {A.~A.}\ \bibnamefont {Kiselev}}, \bibinfo {author}
  {\bibfnamefont {B.~H.}\ \bibnamefont {Fong}}, \bibinfo {author}
  {\bibfnamefont {K.~S.}\ \bibnamefont {Holabird}}, \bibinfo {author}
  {\bibfnamefont {T.~M.}\ \bibnamefont {Hazard}}, \bibinfo {author}
  {\bibfnamefont {B.}~\bibnamefont {Huang}}, \bibinfo {author} {\bibfnamefont
  {P.~W.}\ \bibnamefont {Deelman}}, \bibinfo {author} {\bibfnamefont
  {I.}~\bibnamefont {Milosavljevic}}, \bibinfo {author} {\bibfnamefont {A.~E.}\
  \bibnamefont {Schmitz}}, \bibinfo {author} {\bibfnamefont {R.~S.}\
  \bibnamefont {Ross}}, \bibinfo {author} {\bibfnamefont {M.~F.}\ \bibnamefont
  {Gyure}}, \ and\ \bibinfo {author} {\bibfnamefont {A.~T.}\ \bibnamefont
  {Hunter}},\ }\href {\doibase 10.1126/sciadv.1500214} {\bibfield  {journal}
  {\bibinfo  {journal} {Science Advances}\ }\textbf {\bibinfo {volume} {1}},\
  \bibinfo {pages} {e1500214} (\bibinfo {year} {2015})}\BibitemShut {NoStop}%
\bibitem [{\citenamefont {Kim}\ \emph {et~al.}(2014)\citenamefont {Kim},
  \citenamefont {Shi}, \citenamefont {Simmons}, \citenamefont {Ward},
  \citenamefont {Prance}, \citenamefont {Koh}, \citenamefont {Gamble},
  \citenamefont {Savage}, \citenamefont {Lagally}, \citenamefont {Friesen},
  \citenamefont {Coppersmith},\ and\ \citenamefont {Eriksson}}]{kim2014a}%
  \BibitemOpen
  \bibfield  {author} {\bibinfo {author} {\bibfnamefont {D.}~\bibnamefont
  {Kim}}, \bibinfo {author} {\bibfnamefont {Z.}~\bibnamefont {Shi}}, \bibinfo
  {author} {\bibfnamefont {C.~B.}\ \bibnamefont {Simmons}}, \bibinfo {author}
  {\bibfnamefont {D.~R.}\ \bibnamefont {Ward}}, \bibinfo {author}
  {\bibfnamefont {J.~R.}\ \bibnamefont {Prance}}, \bibinfo {author}
  {\bibfnamefont {T.~S.}\ \bibnamefont {Koh}}, \bibinfo {author} {\bibfnamefont
  {J.~K.}\ \bibnamefont {Gamble}}, \bibinfo {author} {\bibfnamefont {D.~E.}\
  \bibnamefont {Savage}}, \bibinfo {author} {\bibfnamefont {M.~G.}\
  \bibnamefont {Lagally}}, \bibinfo {author} {\bibfnamefont {M.}~\bibnamefont
  {Friesen}}, \bibinfo {author} {\bibfnamefont {S.~N.}\ \bibnamefont
  {Coppersmith}}, \ and\ \bibinfo {author} {\bibfnamefont {M.~A.}\ \bibnamefont
  {Eriksson}},\ }\href {\doibase 10.1038/nature13407} {\bibfield  {journal}
  {\bibinfo  {journal} {Nature}\ }\textbf {\bibinfo {volume} {511}},\ \bibinfo
  {pages} {70} (\bibinfo {year} {2014})}\BibitemShut {NoStop}%
\bibitem [{\citenamefont {Medford}\ \emph {et~al.}(2013)\citenamefont
  {Medford}, \citenamefont {Beil}, \citenamefont {Taylor}, \citenamefont
  {Bartlett}, \citenamefont {Doherty}, \citenamefont {Rashba}, \citenamefont
  {DiVincenzo}, \citenamefont {Lu}, \citenamefont {Gossard},\ and\
  \citenamefont {Marcus}}]{medford2013}%
  \BibitemOpen
  \bibfield  {author} {\bibinfo {author} {\bibfnamefont {J.}~\bibnamefont
  {Medford}}, \bibinfo {author} {\bibfnamefont {J.}~\bibnamefont {Beil}},
  \bibinfo {author} {\bibfnamefont {J.~M.}\ \bibnamefont {Taylor}}, \bibinfo
  {author} {\bibfnamefont {S.~D.}\ \bibnamefont {Bartlett}}, \bibinfo {author}
  {\bibfnamefont {A.~C.}\ \bibnamefont {Doherty}}, \bibinfo {author}
  {\bibfnamefont {E.~I.}\ \bibnamefont {Rashba}}, \bibinfo {author}
  {\bibfnamefont {D.~P.}\ \bibnamefont {DiVincenzo}}, \bibinfo {author}
  {\bibfnamefont {H.}~\bibnamefont {Lu}}, \bibinfo {author} {\bibfnamefont
  {A.~C.}\ \bibnamefont {Gossard}}, \ and\ \bibinfo {author} {\bibfnamefont
  {C.~M.}\ \bibnamefont {Marcus}},\ }\href {\doibase 10.1038/nnano.2013.168}
  {\bibfield  {journal} {\bibinfo  {journal} {Nature Nanotechnology}\ }\textbf
  {\bibinfo {volume} {8}},\ \bibinfo {pages} {654} (\bibinfo {year}
  {2013})}\BibitemShut {NoStop}%
\bibitem [{\citenamefont {Cerfontaine}\ \emph {et~al.}(2016)\citenamefont
  {Cerfontaine}, \citenamefont {Botzem}, \citenamefont {Humpohl}, \citenamefont
  {Schuh}, \citenamefont {Bougeard},\ and\ \citenamefont
  {Bluhm}}]{cerfontaine2016}%
  \BibitemOpen
  \bibfield  {author} {\bibinfo {author} {\bibfnamefont {P.}~\bibnamefont
  {Cerfontaine}}, \bibinfo {author} {\bibfnamefont {T.}~\bibnamefont {Botzem}},
  \bibinfo {author} {\bibfnamefont {S.~S.}\ \bibnamefont {Humpohl}}, \bibinfo
  {author} {\bibfnamefont {D.}~\bibnamefont {Schuh}}, \bibinfo {author}
  {\bibfnamefont {D.}~\bibnamefont {Bougeard}}, \ and\ \bibinfo {author}
  {\bibfnamefont {H.}~\bibnamefont {Bluhm}},\ }\href@noop {} {\bibfield
  {journal} {\bibinfo  {journal} {arXiv:1606.01897 [cond-mat,
  physics:quant-ph]}\ } (\bibinfo {year} {2016})},\ \Eprint
  {http://arxiv.org/abs/1606.01897} {arXiv:1606.01897 [cond-mat,
  physics:quant-ph]} \BibitemShut {NoStop}%
\bibitem [{\citenamefont {Nichol}\ \emph {et~al.}(2017)\citenamefont {Nichol},
  \citenamefont {Orona}, \citenamefont {Harvey}, \citenamefont {Fallahi},
  \citenamefont {Gardner}, \citenamefont {Manfra},\ and\ \citenamefont
  {Yacoby}}]{nichol2017}%
  \BibitemOpen
  \bibfield  {author} {\bibinfo {author} {\bibfnamefont {J.~M.}\ \bibnamefont
  {Nichol}}, \bibinfo {author} {\bibfnamefont {L.~A.}\ \bibnamefont {Orona}},
  \bibinfo {author} {\bibfnamefont {S.~P.}\ \bibnamefont {Harvey}}, \bibinfo
  {author} {\bibfnamefont {S.}~\bibnamefont {Fallahi}}, \bibinfo {author}
  {\bibfnamefont {G.~C.}\ \bibnamefont {Gardner}}, \bibinfo {author}
  {\bibfnamefont {M.~J.}\ \bibnamefont {Manfra}}, \ and\ \bibinfo {author}
  {\bibfnamefont {A.}~\bibnamefont {Yacoby}},\ }\href {\doibase
  10.1038/s41534-016-0003-1} {\bibfield  {journal} {\bibinfo  {journal} {npj
  Quantum Information}\ }\textbf {\bibinfo {volume} {3}},\ \bibinfo {pages} {3}
  (\bibinfo {year} {2017})}\BibitemShut {NoStop}%
\bibitem [{\citenamefont {Shulman}\ \emph {et~al.}(2012)\citenamefont
  {Shulman}, \citenamefont {Dial}, \citenamefont {Harvey}, \citenamefont
  {Bluhm}, \citenamefont {Umansky},\ and\ \citenamefont
  {Yacoby}}]{shulman2012}%
  \BibitemOpen
  \bibfield  {author} {\bibinfo {author} {\bibfnamefont {M.~D.}\ \bibnamefont
  {Shulman}}, \bibinfo {author} {\bibfnamefont {O.~E.}\ \bibnamefont {Dial}},
  \bibinfo {author} {\bibfnamefont {S.~P.}\ \bibnamefont {Harvey}}, \bibinfo
  {author} {\bibfnamefont {H.}~\bibnamefont {Bluhm}}, \bibinfo {author}
  {\bibfnamefont {V.}~\bibnamefont {Umansky}}, \ and\ \bibinfo {author}
  {\bibfnamefont {A.}~\bibnamefont {Yacoby}},\ }\href {\doibase
  10.1126/science.1217692} {\bibfield  {journal} {\bibinfo  {journal}
  {Science}\ }\textbf {\bibinfo {volume} {336}},\ \bibinfo {pages} {202}
  (\bibinfo {year} {2012})}\BibitemShut {NoStop}%
\bibitem [{\citenamefont {Viennot}\ \emph {et~al.}(2015)\citenamefont
  {Viennot}, \citenamefont {Dartiailh}, \citenamefont {Cottet},\ and\
  \citenamefont {Kontos}}]{viennot2015}%
  \BibitemOpen
  \bibfield  {author} {\bibinfo {author} {\bibfnamefont {J.~J.}\ \bibnamefont
  {Viennot}}, \bibinfo {author} {\bibfnamefont {M.~C.}\ \bibnamefont
  {Dartiailh}}, \bibinfo {author} {\bibfnamefont {A.}~\bibnamefont {Cottet}}, \
  and\ \bibinfo {author} {\bibfnamefont {T.}~\bibnamefont {Kontos}},\ }\href
  {\doibase 10.1126/science.aaa3786} {\bibfield  {journal} {\bibinfo  {journal}
  {Science}\ }\textbf {\bibinfo {volume} {349}},\ \bibinfo {pages} {408}
  (\bibinfo {year} {2015})}\BibitemShut {NoStop}%
\bibitem [{\citenamefont {Mi}\ \emph {et~al.}(2018)\citenamefont {Mi},
  \citenamefont {Benito}, \citenamefont {Putz}, \citenamefont {Zajac},
  \citenamefont {Taylor}, \citenamefont {Burkard},\ and\ \citenamefont
  {Petta}}]{mi2018}%
  \BibitemOpen
  \bibfield  {author} {\bibinfo {author} {\bibfnamefont {X.}~\bibnamefont
  {Mi}}, \bibinfo {author} {\bibfnamefont {M.}~\bibnamefont {Benito}}, \bibinfo
  {author} {\bibfnamefont {S.}~\bibnamefont {Putz}}, \bibinfo {author}
  {\bibfnamefont {D.~M.}\ \bibnamefont {Zajac}}, \bibinfo {author}
  {\bibfnamefont {J.~M.}\ \bibnamefont {Taylor}}, \bibinfo {author}
  {\bibfnamefont {G.}~\bibnamefont {Burkard}}, \ and\ \bibinfo {author}
  {\bibfnamefont {J.~R.}\ \bibnamefont {Petta}},\ }\href {\doibase
  10.1038/nature25769} {\bibfield  {journal} {\bibinfo  {journal} {Nature}\ }
  (\bibinfo {year} {2018}),\ 10.1038/nature25769}\BibitemShut {NoStop}%
\bibitem [{\citenamefont {Samkharadze}\ \emph {et~al.}(2018)\citenamefont
  {Samkharadze}, \citenamefont {Zheng}, \citenamefont {Kalhor}, \citenamefont
  {Brousse}, \citenamefont {Sammak}, \citenamefont {Mendes}, \citenamefont
  {Blais}, \citenamefont {Scappucci},\ and\ \citenamefont
  {Vandersypen}}]{samkharadze2018}%
  \BibitemOpen
  \bibfield  {author} {\bibinfo {author} {\bibfnamefont {N.}~\bibnamefont
  {Samkharadze}}, \bibinfo {author} {\bibfnamefont {G.}~\bibnamefont {Zheng}},
  \bibinfo {author} {\bibfnamefont {N.}~\bibnamefont {Kalhor}}, \bibinfo
  {author} {\bibfnamefont {D.}~\bibnamefont {Brousse}}, \bibinfo {author}
  {\bibfnamefont {A.}~\bibnamefont {Sammak}}, \bibinfo {author} {\bibfnamefont
  {U.~C.}\ \bibnamefont {Mendes}}, \bibinfo {author} {\bibfnamefont
  {A.}~\bibnamefont {Blais}}, \bibinfo {author} {\bibfnamefont
  {G.}~\bibnamefont {Scappucci}}, \ and\ \bibinfo {author} {\bibfnamefont
  {L.~M.~K.}\ \bibnamefont {Vandersypen}},\ }\href {\doibase
  10.1126/science.aar4054} {\bibfield  {journal} {\bibinfo  {journal}
  {Science}\ ,\ \bibinfo {pages} {eaar4054}} (\bibinfo {year}
  {2018})}\BibitemShut {NoStop}%
\bibitem [{\citenamefont {Landig}\ \emph {et~al.}(2017)\citenamefont {Landig},
  \citenamefont {Koski}, \citenamefont {Scarlino}, \citenamefont {Mendes},
  \citenamefont {Blais}, \citenamefont {Reichl}, \citenamefont {Wegscheider},
  \citenamefont {Wallraff}, \citenamefont {Ensslin},\ and\ \citenamefont
  {Ihn}}]{landig2017}%
  \BibitemOpen
  \bibfield  {author} {\bibinfo {author} {\bibfnamefont {A.~J.}\ \bibnamefont
  {Landig}}, \bibinfo {author} {\bibfnamefont {J.~V.}\ \bibnamefont {Koski}},
  \bibinfo {author} {\bibfnamefont {P.}~\bibnamefont {Scarlino}}, \bibinfo
  {author} {\bibfnamefont {U.~C.}\ \bibnamefont {Mendes}}, \bibinfo {author}
  {\bibfnamefont {A.}~\bibnamefont {Blais}}, \bibinfo {author} {\bibfnamefont
  {C.}~\bibnamefont {Reichl}}, \bibinfo {author} {\bibfnamefont
  {W.}~\bibnamefont {Wegscheider}}, \bibinfo {author} {\bibfnamefont
  {A.}~\bibnamefont {Wallraff}}, \bibinfo {author} {\bibfnamefont
  {K.}~\bibnamefont {Ensslin}}, \ and\ \bibinfo {author} {\bibfnamefont
  {T.}~\bibnamefont {Ihn}},\ }\href@noop {} {\bibfield  {journal} {\bibinfo
  {journal} {arXiv:1711.01932 [cond-mat]}\ } (\bibinfo {year} {2017})},\
  \Eprint {http://arxiv.org/abs/1711.01932} {arXiv:1711.01932 [cond-mat]}
  \BibitemShut {NoStop}%
\bibitem [{\citenamefont {Blais}\ \emph {et~al.}(2004)\citenamefont {Blais},
  \citenamefont {Huang}, \citenamefont {Wallraff}, \citenamefont {Girvin},\
  and\ \citenamefont {Schoelkopf}}]{blais2004}%
  \BibitemOpen
  \bibfield  {author} {\bibinfo {author} {\bibfnamefont {A.}~\bibnamefont
  {Blais}}, \bibinfo {author} {\bibfnamefont {R.-S.}\ \bibnamefont {Huang}},
  \bibinfo {author} {\bibfnamefont {A.}~\bibnamefont {Wallraff}}, \bibinfo
  {author} {\bibfnamefont {S.}~\bibnamefont {Girvin}}, \ and\ \bibinfo {author}
  {\bibfnamefont {R.}~\bibnamefont {Schoelkopf}},\ }\href {\doibase
  10.1103/PhysRevA.69.062320} {\bibfield  {journal} {\bibinfo  {journal}
  {Physical Review A}\ }\textbf {\bibinfo {volume} {69}} (\bibinfo {year}
  {2004}),\ 10.1103/PhysRevA.69.062320}\BibitemShut {NoStop}%
\bibitem [{\citenamefont {Kerman}(2013)}]{kerman2013}%
  \BibitemOpen
  \bibfield  {author} {\bibinfo {author} {\bibfnamefont {A.~J.}\ \bibnamefont
  {Kerman}},\ }\href {\doibase 10.1088/1367-2630/15/12/123011} {\bibfield
  {journal} {\bibinfo  {journal} {New Journal of Physics}\ }\textbf {\bibinfo
  {volume} {15}},\ \bibinfo {pages} {123011} (\bibinfo {year}
  {2013})}\BibitemShut {NoStop}%
\bibitem [{\citenamefont {Schuetz}\ \emph {et~al.}(2017)\citenamefont
  {Schuetz}, \citenamefont {Giedke}, \citenamefont {Vandersypen},\ and\
  \citenamefont {Cirac}}]{schuetz2017}%
  \BibitemOpen
  \bibfield  {author} {\bibinfo {author} {\bibfnamefont {M.~J.~A.}\
  \bibnamefont {Schuetz}}, \bibinfo {author} {\bibfnamefont {G.}~\bibnamefont
  {Giedke}}, \bibinfo {author} {\bibfnamefont {L.~M.~K.}\ \bibnamefont
  {Vandersypen}}, \ and\ \bibinfo {author} {\bibfnamefont {J.~I.}\ \bibnamefont
  {Cirac}},\ }\href {\doibase 10.1103/PhysRevA.95.052335} {\bibfield  {journal}
  {\bibinfo  {journal} {Physical Review A}\ }\textbf {\bibinfo {volume} {95}},\
  \bibinfo {pages} {052335} (\bibinfo {year} {2017})}\BibitemShut {NoStop}%
\bibitem [{\citenamefont {Billangeon}\ \emph {et~al.}(2015)\citenamefont
  {Billangeon}, \citenamefont {Tsai},\ and\ \citenamefont
  {Nakamura}}]{billangeon2015}%
  \BibitemOpen
  \bibfield  {author} {\bibinfo {author} {\bibfnamefont {P.-M.}\ \bibnamefont
  {Billangeon}}, \bibinfo {author} {\bibfnamefont {J.~S.}\ \bibnamefont
  {Tsai}}, \ and\ \bibinfo {author} {\bibfnamefont {Y.}~\bibnamefont
  {Nakamura}},\ }\href {\doibase 10.1103/PhysRevB.91.094517} {\bibfield
  {journal} {\bibinfo  {journal} {Physical Review B}\ }\textbf {\bibinfo
  {volume} {91}},\ \bibinfo {pages} {094517} (\bibinfo {year}
  {2015})}\BibitemShut {NoStop}%
\bibitem [{\citenamefont {Didier}\ \emph {et~al.}(2015)\citenamefont {Didier},
  \citenamefont {Bourassa},\ and\ \citenamefont {Blais}}]{didier2015}%
  \BibitemOpen
  \bibfield  {author} {\bibinfo {author} {\bibfnamefont {N.}~\bibnamefont
  {Didier}}, \bibinfo {author} {\bibfnamefont {J.}~\bibnamefont {Bourassa}}, \
  and\ \bibinfo {author} {\bibfnamefont {A.}~\bibnamefont {Blais}},\ }\href
  {\doibase 10.1103/PhysRevLett.115.203601} {\bibfield  {journal} {\bibinfo
  {journal} {Physical Review Letters}\ }\textbf {\bibinfo {volume} {115}},\
  \bibinfo {pages} {203601} (\bibinfo {year} {2015})}\BibitemShut {NoStop}%
\bibitem [{\citenamefont {Richer}\ and\ \citenamefont
  {DiVincenzo}(2016)}]{richer2016}%
  \BibitemOpen
  \bibfield  {author} {\bibinfo {author} {\bibfnamefont {S.}~\bibnamefont
  {Richer}}\ and\ \bibinfo {author} {\bibfnamefont {D.}~\bibnamefont
  {DiVincenzo}},\ }\href {\doibase 10.1103/PhysRevB.93.134501} {\bibfield
  {journal} {\bibinfo  {journal} {Physical Review B}\ }\textbf {\bibinfo
  {volume} {93}},\ \bibinfo {pages} {134501} (\bibinfo {year}
  {2016})}\BibitemShut {NoStop}%
\bibitem [{\citenamefont {Royer}\ \emph {et~al.}(2017)\citenamefont {Royer},
  \citenamefont {Grimsmo}, \citenamefont {Didier},\ and\ \citenamefont
  {Blais}}]{royer2017}%
  \BibitemOpen
  \bibfield  {author} {\bibinfo {author} {\bibfnamefont {B.}~\bibnamefont
  {Royer}}, \bibinfo {author} {\bibfnamefont {A.~L.}\ \bibnamefont {Grimsmo}},
  \bibinfo {author} {\bibfnamefont {N.}~\bibnamefont {Didier}}, \ and\ \bibinfo
  {author} {\bibfnamefont {A.}~\bibnamefont {Blais}},\ }\href {\doibase
  10.22331/q-2017-05-11-11} {\bibfield  {journal} {\bibinfo  {journal}
  {Quantum}\ }\textbf {\bibinfo {volume} {1}},\ \bibinfo {pages} {11} (\bibinfo
  {year} {2017})}\BibitemShut {NoStop}%
\bibitem [{\citenamefont {Elman}\ \emph {et~al.}(2017)\citenamefont {Elman},
  \citenamefont {Bartlett},\ and\ \citenamefont {Doherty}}]{elman2017a}%
  \BibitemOpen
  \bibfield  {author} {\bibinfo {author} {\bibfnamefont {S.~J.}\ \bibnamefont
  {Elman}}, \bibinfo {author} {\bibfnamefont {S.~D.}\ \bibnamefont {Bartlett}},
  \ and\ \bibinfo {author} {\bibfnamefont {A.~C.}\ \bibnamefont {Doherty}},\
  }\href {\doibase 10.1103/PhysRevB.96.115407} {\bibfield  {journal} {\bibinfo
  {journal} {Physical Review B}\ }\textbf {\bibinfo {volume} {96}},\ \bibinfo
  {pages} {115407} (\bibinfo {year} {2017})}\BibitemShut {NoStop}%
\bibitem [{\citenamefont {Gambetta}\ \emph {et~al.}(2008)\citenamefont
  {Gambetta}, \citenamefont {Blais}, \citenamefont {Boissonneault},
  \citenamefont {Houck}, \citenamefont {Schuster},\ and\ \citenamefont
  {Girvin}}]{gambetta2008}%
  \BibitemOpen
  \bibfield  {author} {\bibinfo {author} {\bibfnamefont {J.}~\bibnamefont
  {Gambetta}}, \bibinfo {author} {\bibfnamefont {A.}~\bibnamefont {Blais}},
  \bibinfo {author} {\bibfnamefont {M.}~\bibnamefont {Boissonneault}}, \bibinfo
  {author} {\bibfnamefont {A.~A.}\ \bibnamefont {Houck}}, \bibinfo {author}
  {\bibfnamefont {D.~I.}\ \bibnamefont {Schuster}}, \ and\ \bibinfo {author}
  {\bibfnamefont {S.~M.}\ \bibnamefont {Girvin}},\ }\href {\doibase
  10.1103/PhysRevA.77.012112} {\bibfield  {journal} {\bibinfo  {journal}
  {Physical Review A}\ }\textbf {\bibinfo {volume} {77}},\ \bibinfo {pages}
  {012112} (\bibinfo {year} {2008})}\BibitemShut {NoStop}%
\bibitem [{\citenamefont {Leibfried}\ \emph {et~al.}(2003)\citenamefont
  {Leibfried}, \citenamefont {Blatt}, \citenamefont {Monroe},\ and\
  \citenamefont {Wineland}}]{leibfried2003a}%
  \BibitemOpen
  \bibfield  {author} {\bibinfo {author} {\bibfnamefont {D.}~\bibnamefont
  {Leibfried}}, \bibinfo {author} {\bibfnamefont {R.}~\bibnamefont {Blatt}},
  \bibinfo {author} {\bibfnamefont {C.}~\bibnamefont {Monroe}}, \ and\ \bibinfo
  {author} {\bibfnamefont {D.}~\bibnamefont {Wineland}},\ }\href@noop {}
  {\bibfield  {journal} {\bibinfo  {journal} {Reviews of Modern Physics}\
  }\textbf {\bibinfo {volume} {75}},\ \bibinfo {pages} {281} (\bibinfo {year}
  {2003})}\BibitemShut {NoStop}%
\bibitem [{\citenamefont {Leung}\ \emph {et~al.}(2018)\citenamefont {Leung},
  \citenamefont {Landsman}, \citenamefont {Figgatt}, \citenamefont {Linke},
  \citenamefont {Monroe},\ and\ \citenamefont {Brown}}]{leung2018}%
  \BibitemOpen
  \bibfield  {author} {\bibinfo {author} {\bibfnamefont {P.~H.}\ \bibnamefont
  {Leung}}, \bibinfo {author} {\bibfnamefont {K.~A.}\ \bibnamefont {Landsman}},
  \bibinfo {author} {\bibfnamefont {C.}~\bibnamefont {Figgatt}}, \bibinfo
  {author} {\bibfnamefont {N.~M.}\ \bibnamefont {Linke}}, \bibinfo {author}
  {\bibfnamefont {C.}~\bibnamefont {Monroe}}, \ and\ \bibinfo {author}
  {\bibfnamefont {K.~R.}\ \bibnamefont {Brown}},\ }\href {\doibase
  10.1103/PhysRevLett.120.020501} {\bibfield  {journal} {\bibinfo  {journal}
  {Physical Review Letters}\ }\textbf {\bibinfo {volume} {120}},\ \bibinfo
  {pages} {020501} (\bibinfo {year} {2018})}\BibitemShut {NoStop}%
\bibitem [{\citenamefont {Roos}(2008)}]{roos2008}%
  \BibitemOpen
  \bibfield  {author} {\bibinfo {author} {\bibfnamefont {C.~F.}\ \bibnamefont
  {Roos}},\ }\href {\doibase 10.1088/1367-2630/10/1/013002} {\bibfield
  {journal} {\bibinfo  {journal} {New Journal of Physics}\ }\textbf {\bibinfo
  {volume} {10}},\ \bibinfo {pages} {013002} (\bibinfo {year}
  {2008})}\BibitemShut {NoStop}%
\bibitem [{\citenamefont {Stockklauser}\ \emph {et~al.}(2017)\citenamefont
  {Stockklauser}, \citenamefont {Scarlino}, \citenamefont {Koski},
  \citenamefont {Gasparinetti}, \citenamefont {Andersen}, \citenamefont
  {Reichl}, \citenamefont {Wegscheider}, \citenamefont {Ihn}, \citenamefont
  {Ensslin},\ and\ \citenamefont {Wallraff}}]{stockklauser2017}%
  \BibitemOpen
  \bibfield  {author} {\bibinfo {author} {\bibfnamefont {A.}~\bibnamefont
  {Stockklauser}}, \bibinfo {author} {\bibfnamefont {P.}~\bibnamefont
  {Scarlino}}, \bibinfo {author} {\bibfnamefont {J.}~\bibnamefont {Koski}},
  \bibinfo {author} {\bibfnamefont {S.}~\bibnamefont {Gasparinetti}}, \bibinfo
  {author} {\bibfnamefont {C.~K.}\ \bibnamefont {Andersen}}, \bibinfo {author}
  {\bibfnamefont {C.}~\bibnamefont {Reichl}}, \bibinfo {author} {\bibfnamefont
  {W.}~\bibnamefont {Wegscheider}}, \bibinfo {author} {\bibfnamefont
  {T.}~\bibnamefont {Ihn}}, \bibinfo {author} {\bibfnamefont {K.}~\bibnamefont
  {Ensslin}}, \ and\ \bibinfo {author} {\bibfnamefont {A.}~\bibnamefont
  {Wallraff}},\ }\href {\doibase 10.1103/PhysRevX.7.011030} {\bibfield
  {journal} {\bibinfo  {journal} {Physical Review X}\ }\textbf {\bibinfo
  {volume} {7}} (\bibinfo {year} {2017}),\ 10.1103/PhysRevX.7.011030},\ \Eprint
  {http://arxiv.org/abs/1701.03433} {arXiv:1701.03433} \BibitemShut {NoStop}%
\bibitem [{\citenamefont {Samkharadze}\ \emph {et~al.}(2016)\citenamefont
  {Samkharadze}, \citenamefont {Bruno}, \citenamefont {Scarlino}, \citenamefont
  {Zheng}, \citenamefont {DiVincenzo}, \citenamefont {DiCarlo},\ and\
  \citenamefont {Vandersypen}}]{samkharadze2016}%
  \BibitemOpen
  \bibfield  {author} {\bibinfo {author} {\bibfnamefont {N.}~\bibnamefont
  {Samkharadze}}, \bibinfo {author} {\bibfnamefont {A.}~\bibnamefont {Bruno}},
  \bibinfo {author} {\bibfnamefont {P.}~\bibnamefont {Scarlino}}, \bibinfo
  {author} {\bibfnamefont {G.}~\bibnamefont {Zheng}}, \bibinfo {author}
  {\bibfnamefont {D.~P.}\ \bibnamefont {DiVincenzo}}, \bibinfo {author}
  {\bibfnamefont {L.}~\bibnamefont {DiCarlo}}, \ and\ \bibinfo {author}
  {\bibfnamefont {L.~M.~K.}\ \bibnamefont {Vandersypen}},\ }\href {\doibase
  10.1103/PhysRevApplied.5.044004} {\bibfield  {journal} {\bibinfo  {journal}
  {Physical Review Applied}\ }\textbf {\bibinfo {volume} {5}},\ \bibinfo
  {pages} {044004} (\bibinfo {year} {2016})}\BibitemShut {NoStop}%
\bibitem [{\citenamefont {{van der Wiel}}\ \emph {et~al.}(2002)\citenamefont
  {{van der Wiel}}, \citenamefont {De~Franceschi}, \citenamefont {Elzerman},
  \citenamefont {Fujisawa}, \citenamefont {Tarucha},\ and\ \citenamefont
  {Kouwenhoven}}]{vanderwiel2002}%
  \BibitemOpen
  \bibfield  {author} {\bibinfo {author} {\bibfnamefont {W.~G.}\ \bibnamefont
  {{van der Wiel}}}, \bibinfo {author} {\bibfnamefont {S.}~\bibnamefont
  {De~Franceschi}}, \bibinfo {author} {\bibfnamefont {J.~M.}\ \bibnamefont
  {Elzerman}}, \bibinfo {author} {\bibfnamefont {T.}~\bibnamefont {Fujisawa}},
  \bibinfo {author} {\bibfnamefont {S.}~\bibnamefont {Tarucha}}, \ and\
  \bibinfo {author} {\bibfnamefont {L.~P.}\ \bibnamefont {Kouwenhoven}},\
  }\href {\doibase 10.1103/RevModPhys.75.1} {\bibfield  {journal} {\bibinfo
  {journal} {Reviews of Modern Physics}\ }\textbf {\bibinfo {volume} {75}},\
  \bibinfo {pages} {1} (\bibinfo {year} {2002})}\BibitemShut {NoStop}%
\bibitem [{\citenamefont {Poyatos}\ \emph {et~al.}(1997)\citenamefont
  {Poyatos}, \citenamefont {Cirac},\ and\ \citenamefont
  {Zoller}}]{poyatos1997}%
  \BibitemOpen
  \bibfield  {author} {\bibinfo {author} {\bibfnamefont {J.~F.}\ \bibnamefont
  {Poyatos}}, \bibinfo {author} {\bibfnamefont {J.~I.}\ \bibnamefont {Cirac}},
  \ and\ \bibinfo {author} {\bibfnamefont {P.}~\bibnamefont {Zoller}},\ }\href
  {\doibase 10.1103/PhysRevLett.78.390} {\bibfield  {journal} {\bibinfo
  {journal} {Physical Review Letters}\ }\textbf {\bibinfo {volume} {78}},\
  \bibinfo {pages} {390} (\bibinfo {year} {1997})}\BibitemShut {NoStop}%
\bibitem [{\citenamefont {Nielsen}(2002)}]{nielsen2002}%
  \BibitemOpen
  \bibfield  {author} {\bibinfo {author} {\bibfnamefont {M.~A.}\ \bibnamefont
  {Nielsen}},\ }\href {\doibase 10.1016/S0375-9601(02)01272-0} {\bibfield
  {journal} {\bibinfo  {journal} {Physics Letters A}\ }\textbf {\bibinfo
  {volume} {303}},\ \bibinfo {pages} {249} (\bibinfo {year}
  {2002})}\BibitemShut {NoStop}%
\bibitem [{\citenamefont {Dial}\ \emph {et~al.}(2013)\citenamefont {Dial},
  \citenamefont {Shulman}, \citenamefont {Harvey}, \citenamefont {Bluhm},
  \citenamefont {Umansky},\ and\ \citenamefont {Yacoby}}]{dial2013}%
  \BibitemOpen
  \bibfield  {author} {\bibinfo {author} {\bibfnamefont {O.}~\bibnamefont
  {Dial}}, \bibinfo {author} {\bibfnamefont {M.}~\bibnamefont {Shulman}},
  \bibinfo {author} {\bibfnamefont {S.}~\bibnamefont {Harvey}}, \bibinfo
  {author} {\bibfnamefont {H.}~\bibnamefont {Bluhm}}, \bibinfo {author}
  {\bibfnamefont {V.}~\bibnamefont {Umansky}}, \ and\ \bibinfo {author}
  {\bibfnamefont {A.}~\bibnamefont {Yacoby}},\ }\href {\doibase
  10.1103/PhysRevLett.110.146804} {\bibfield  {journal} {\bibinfo  {journal}
  {Physical Review Letters}\ }\textbf {\bibinfo {volume} {110}} (\bibinfo
  {year} {2013}),\ 10.1103/PhysRevLett.110.146804}\BibitemShut {NoStop}%
\bibitem [{\citenamefont {Cywi{\'n}ski}\ \emph {et~al.}(2008)\citenamefont
  {Cywi{\'n}ski}, \citenamefont {Lutchyn}, \citenamefont {Nave},\ and\
  \citenamefont {Das~Sarma}}]{cywinski2008}%
  \BibitemOpen
  \bibfield  {author} {\bibinfo {author} {\bibfnamefont {{\L}.}~\bibnamefont
  {Cywi{\'n}ski}}, \bibinfo {author} {\bibfnamefont {R.}~\bibnamefont
  {Lutchyn}}, \bibinfo {author} {\bibfnamefont {C.}~\bibnamefont {Nave}}, \
  and\ \bibinfo {author} {\bibfnamefont {S.}~\bibnamefont {Das~Sarma}},\ }\href
  {\doibase 10.1103/PhysRevB.77.174509} {\bibfield  {journal} {\bibinfo
  {journal} {Physical Review B}\ }\textbf {\bibinfo {volume} {77}} (\bibinfo
  {year} {2008}),\ 10.1103/PhysRevB.77.174509}\BibitemShut {NoStop}%
\bibitem [{\citenamefont {Kogan}(2008)}]{kogan2008}%
  \BibitemOpen
  \bibfield  {author} {\bibinfo {author} {\bibfnamefont {S.}~\bibnamefont
  {Kogan}},\ }\href@noop {} {\emph {\bibinfo {title} {Electronic {{Noise}} and
  {{Fluctuations}} in {{Solids}}}}}\ (\bibinfo  {publisher} {{Cambridge
  University Press}},\ \bibinfo {year} {2008})\BibitemShut {NoStop}%
\bibitem [{\citenamefont {Bell}\ \emph {et~al.}(2012)\citenamefont {Bell},
  \citenamefont {Sadovskyy}, \citenamefont {Ioffe}, \citenamefont {Kitaev},\
  and\ \citenamefont {Gershenson}}]{bell2012a}%
  \BibitemOpen
  \bibfield  {author} {\bibinfo {author} {\bibfnamefont {M.~T.}\ \bibnamefont
  {Bell}}, \bibinfo {author} {\bibfnamefont {I.~A.}\ \bibnamefont {Sadovskyy}},
  \bibinfo {author} {\bibfnamefont {L.~B.}\ \bibnamefont {Ioffe}}, \bibinfo
  {author} {\bibfnamefont {A.~Y.}\ \bibnamefont {Kitaev}}, \ and\ \bibinfo
  {author} {\bibfnamefont {M.~E.}\ \bibnamefont {Gershenson}},\ }\href
  {\doibase 10.1103/PhysRevLett.109.137003} {\bibfield  {journal} {\bibinfo
  {journal} {Physical Review Letters}\ }\textbf {\bibinfo {volume} {109}}
  (\bibinfo {year} {2012}),\ 10.1103/PhysRevLett.109.137003}\BibitemShut
  {NoStop}%
\bibitem [{\citenamefont {Mi}\ \emph {et~al.}(2017)\citenamefont {Mi},
  \citenamefont {Cady}, \citenamefont {Zajac}, \citenamefont {Stehlik},
  \citenamefont {Edge},\ and\ \citenamefont {Petta}}]{mi2017}%
  \BibitemOpen
  \bibfield  {author} {\bibinfo {author} {\bibfnamefont {X.}~\bibnamefont
  {Mi}}, \bibinfo {author} {\bibfnamefont {J.~V.}\ \bibnamefont {Cady}},
  \bibinfo {author} {\bibfnamefont {D.~M.}\ \bibnamefont {Zajac}}, \bibinfo
  {author} {\bibfnamefont {J.}~\bibnamefont {Stehlik}}, \bibinfo {author}
  {\bibfnamefont {L.~F.}\ \bibnamefont {Edge}}, \ and\ \bibinfo {author}
  {\bibfnamefont {J.~R.}\ \bibnamefont {Petta}},\ }\href {\doibase
  10.1063/1.4974536} {\bibfield  {journal} {\bibinfo  {journal} {Applied
  Physics Letters}\ }\textbf {\bibinfo {volume} {110}},\ \bibinfo {pages}
  {043502} (\bibinfo {year} {2017})}\BibitemShut {NoStop}%
\bibitem [{\citenamefont {Leibfried}(2004)}]{leibfried2004}%
  \BibitemOpen
  \bibfield  {author} {\bibinfo {author} {\bibfnamefont {D.}~\bibnamefont
  {Leibfried}},\ }\href {\doibase 10.1126/science.1097576} {\bibfield
  {journal} {\bibinfo  {journal} {Science}\ }\textbf {\bibinfo {volume}
  {304}},\ \bibinfo {pages} {1476} (\bibinfo {year} {2004})}\BibitemShut
  {NoStop}%
\bibitem [{\citenamefont {Wiseman}\ and\ \citenamefont
  {Milburn}(2010)}]{wiseman2010}%
  \BibitemOpen
  \bibfield  {author} {\bibinfo {author} {\bibfnamefont {H.~M.}\ \bibnamefont
  {Wiseman}}\ and\ \bibinfo {author} {\bibfnamefont {G.~J.}\ \bibnamefont
  {Milburn}},\ }\href@noop {} {\emph {\bibinfo {title} {Quantum {{Measurement}}
  and {{Control}}}}}\ (\bibinfo  {publisher} {{Cambridge University Press}},\
  \bibinfo {year} {2010})\BibitemShut {NoStop}%
\bibitem [{\citenamefont {Carmichael}(1993)}]{carmichael1993}%
  \BibitemOpen
  \bibfield  {author} {\bibinfo {author} {\bibfnamefont {H.}~\bibnamefont
  {Carmichael}},\ }\href@noop {} {\emph {\bibinfo {title} {An {{Open Systems
  Approach}} to {{Quantum Optics}}: {{Lectures Presented}} at the
  {{Universit{\'e} Libre}} de {{Bruxelles}}, {{October}} 28 to {{November}} 4,
  1991}}},\ Lecture Notes in Physics Monographs\ (\bibinfo  {publisher}
  {{Springer-Verlag}},\ \bibinfo {address} {Berlin Heidelberg},\ \bibinfo
  {year} {1993})\BibitemShut {NoStop}%
\end{thebibliography}%

\end{document}